\documentclass[english,pra,preprint,nofootinbib,superscriptaddress]{revtex4-1}
\usepackage[T1]{fontenc}
\usepackage[latin9]{inputenc}
\usepackage{geometry}
\usepackage{color}
\usepackage{babel}
\usepackage{amsmath}
\usepackage{amsthm}
\usepackage{amssymb}
\usepackage{graphicx}
\usepackage{esint}
\usepackage{braket}
\PassOptionsToPackage{normalem}{ulem}
\usepackage{ulem}
\usepackage[unicode=true,pdfusetitle,
 bookmarks=true,bookmarksnumbered=false,bookmarksopen=false,
 breaklinks=false,pdfborder={0 0 0},backref=false,colorlinks=true]
 {hyperref}

\begin{document}
\global\long\def\x{\times}
\global\long\def\t{\cdot}
\global\long\def\d{\mathrm{d}}
\global\long\def\ket#1{\left|#1\right\rangle }
\global\long\def\bra#1{\left\langle #1\right|}
\global\long\def\braket#1#2{\langle#1|#2\rangle}
\global\long\def\braoket#1#2#3{\left\langle #1\middle\vert#2\middle\vert#3\right\rangle }
\global\long\def\i{\mathrm{i}}
\global\long\def\e{\mathrm{e}}

\title{Generalized perspective on chiral measurements without magnetic interactions}

\author{Andres F. Ordonez}
\email{ordonez@mbi-berlin.de}
\affiliation{Max-Born-Institut, Berlin, Germany}
\affiliation{Technische Universit\"at Berlin, Berlin, Germany}
\author{Olga Smirnova}
\email{smirnova@mbi-berlin.de}
\affiliation{Max-Born-Institut, Berlin, Germany}
\affiliation{Technische Universit\"at Berlin, Berlin, Germany}

\begin{abstract}
 We present a unified description of several
methods of chiral discrimination based exclusively on electric-dipole interactions.
It includes photoelectron circular dichroism (PECD), enantio-sensitive microwave spectroscopy (EMWS), photoexcitation circular dichroism (PXCD) and
photoelectron-photoexcitation circular dichroism (PXECD). We show
that, in spite of the fact that the physics underlying the appearance
of a chiral response is very different in all these methods, the enantio-sensitive and dichroic
observable in all cases has a unique form. It is a polar vector given by the product of (i) a molecular pseudoscalar and (ii) a field pseudovector specified by the configuration of the electric fields interacting with the isotropic ensemble of chiral molecules. 
The molecular pseudoscalar is a rotationally invariant property, which is composed from different molecule-specific vectors and in the simplest case is a triple product of such vectors. The key property that enables the chiral response is the non-coplanarity of the vectors forming
such triple product. The key property that enables  chiral detection without relying on the chirality of the electromagnetic fields 
is the vectorial nature of the enantio-sensitive observable. Our compact and general expression for this observable shows what
ultimately determines the efficiency of the chiral signal and 
if, or when, it can reach 100\%. 
We also discuss 
the differences between the two phenomena, which rely on the bound states, PXCD and EMWS, and the two phenomena using the continuum states, PECD and PXECD. Finally, we extend these methods to arbitrary polarizations of the electric fields used to induce and probe the chiral response.
\end{abstract}
\maketitle

\section{Introduction}
Right- and left-handed helices are typical examples of 
chiral objects; each
of them cannot be superimposed on its own mirror image. Some molecules
possess the same property; left-handed and right-handed molecules
are called enantiomers. Distinguishing left and right enantiomers
is both vital and difficult \cite{boesl_2016, janssen_2017, chiral_drugs}. Since the XIX century, the helix of circularly polarized light was used to distinguish the two enantiomers of a
chiral molecule, relying on the relatively weak interaction with the magnetic
field as a key mechanism for chiral discrimination. However, in this case the chiral signal\footnote{When referring to the measured signal, we will use the adjective \emph{chiral} as a shorthand for \emph{enantio-sensitive and dichroic.}} is proportional to the ratio of the molecular size 
to the pitch of the light helix, i.e. its
wavelength, generally leading to weak signals in the infrared, visible, and UV regions. 

One can overcome this unfavorable scaling and obtain significantly
higher circular dichroism, at the level of a few percent, in 
several ways. Firstly, one can rely on  
using a strong laser field to enhance the magnetic-dipole transitions 
and interfere them against the electric-dipole ones, as done in chiral high harmonic generation
\cite{cireasa_probing_2015,smirnova_opportunities_2015,ayuso_2018,ayuso2018strong}.
Secondly, one can decrease the pitch of the light helix by using
XUV/X-ray light \cite{zhang2017x,rouxel2017photoinduced}. Yet, in both cases the chiral signal would be equal to  zero within the electric-dipole approximation.

Thus, the discovery of  approaches relying exclusively on
electronic dipole transitions \cite{ritchie_theory_1976,cherepkov1982circular,powis_photoelectron_2000,bowering_asymmetry_2001,fischer_three-wave_2000,fischer2001isotropic,patterson_enantiomer-specific_2013,yurchenko_2016, beaulieu_PXCD_2016, *beaulieu_PXCD}
and yielding a very  high chiral response already in the electric-dipole approximation is both intriguing and beneficial. These techniques include photoelectron circular
dichroism (PECD) \cite{ritchie_theory_1976,cherepkov1982circular,powis_photoelectron_2000,bowering_asymmetry_2001},
enantio-sensitive microwave spectroscopy (EMWS) \cite{patterson_enantiomer-specific_2013, patterson_sensitive_2013, Laane_2018},
photoexcitation circular dichroism (PXCD) \cite{beaulieu_PXCD_2016, *beaulieu_PXCD},
and photoexcitation-photoelectron circular dichroism (PXECD) \cite{beaulieu_PXCD_2016, *beaulieu_PXCD}. 

This new generation of chiral methods leads to very high signals,
up to tens of percent in PECD, which is several orders of magnitude
higher than in standard techniques relying on  magnetic interactions.
Here we present a unified description of several of these methods
working in the perturbative one- and two-photon regimes of the light-molecule
interaction. Results for the multiphoton \cite{lux_circular_2012,lehmann_imaging_2013,lux_photoelectron_2015}
and the strong-field regime \cite{beaulieu_universality_2016, lein_2014} of
PECD will be presented elsewhere. 

We derive a common general formulation for the chiral response encompassing PECD, EMWS, PXCD, and PXECD.
This formulation is based on understanding that these electric-dipole based techniques using non-chiral fields are only possible thanks to vectorial observables. Readers familiar with chiral measurements might be uncomfortable with such statement. Indeed, it is well known that chiral observables are pseudoscalars, not polar vectors. Section \ref{sec:chirality} addresses this issue and describes the role of the lab setup in enantio-sensitive techniques with non-chiral fields. 
In Sec. \ref{sec:Symmetry} we describe how symmetry enforces enantio-sensitivity and dichroism on polar vectors resulting from the electric-dipole interaction. 
Section \ref{sec:vectorial_formulation}  consists of four parts which specify how the information about the handedness of the lab setup and that of the molecular enantiomer  can be decoupled and defined in a common way for the four perturbative dipole  techniques: PECD, PXCD, EMWS, and PXECD. Section \ref{sec:Conclusions} summarizes the conclusions of this work. We  use atomic units throughout the paper.

\section{Chiral measurements and enantio-sensitive observables \label{sec:chirality}}

\begin{figure}
\noindent \begin{centering}
\includegraphics[scale=1.3]{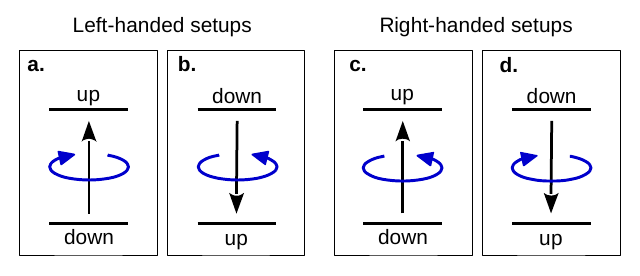}
\par\end{centering}

\caption{The combination of circularly polarized light (blue curved arrows) and a detector (horizontal lines) defining a vector perpendicular to the polarization plane (black vertical arrows) make up a chiral setup. Four possible realizations of such setup are shown. 
Setups \textbf{a} and \textbf{b} are left handed and setups \textbf{c} and \textbf{d} are right handed. 
For a fixed molecular enantiomer (not shown in the figure) setups with the same handedness yield the same result, while setups with opposite handedness yield opposite results.
\label{fig:chiral_setup}}
\end{figure}

The goal of our work is to demonstrate the general concept underlying several chiral measurements which do not use magnetic interactions. Achieving this goal requires two things. First, one should provide a general concept, i.e. address the question ``\emph{what is the key difference between the chiral measurements involving the magnetic component of the light field and those relying only on the electric-dipole approximation?}''. We outline such concept in this section.  Second, 
one should formalize this concept by deriving compact expressions for observables pertinent to the four different experimental setups and establishing connections between them. Such derivations will be presented in   Sec. \ref{sec:vectorial_formulation}.

It is well known that any enantio-sensitive observable should  be a pseudoscalar. However, detectors in any experimental setup measure clicks. Clicks are scalars. Where is the pseudoscalar in a click? 

Let us start with the conventional concept. It is  well known that the handedness of chiral objects can only be probed via interaction with another chiral object, in other words, it is well known that one always needs a chiral reagent to discriminate between opposite enantiomers. A chiral reagent interacts differently with left and right enantiomers.  The chiral reagent can be simply another chiral molecule or chiral light.   Consider, for example, absorption circular dichroism.  Absorption of circularly polarized light by a chiral molecule is the outcome of such an experiment, and this absorption must be different for right and left enantiomers. The difference in absorption is a scalar, however this scalar is just a product of two pseudoscalars, one from the molecule and the other from light. In this particular case the second pseudoscalar is the light helicity (see Appendix \ref{sec:AppendixCD}), which encodes the handedness of the helix traced by the circularly polarized light in space. Thus, we use the chiral probe (chiral reagent) to ``hide'' a molecular pseudoscalar inside a scalar. The molecular pseudoscalar in absorption circular dichroism, as it is well known, is given by the scalar product of electric-dipole and magnetic-dipole vectors. The overall signal is small because the magnetic field interacts very weakly with molecules.

We now turn to methods which do not rely on the interaction with the magnetic component of the light field such as e.g. PECD. 
In PECD the photoionization of an isotropic molecular ensemble with circularly polarized light yields a net photoelectron current in the direction perpendicular to the plane of polarization. The direction of this current can be flipped by either swapping the molecular handedness or the direction of rotation of the field. It is a purely electric-dipole effect: light chirality is not needed at all, i.e. the magnetic field of the incident laser pulse is not used. Thus, we do not use the chiral property of light, yet the chiral signal is very strong. Where is our chiral reagent if the light chirality is not used? The combination of circularly polarized light and a detector that distinguishes the two opposite directions perpendicular to the polarization plane defines a chiral setup (see Fig. \ref{fig:chiral_setup}) whose handedness (a pseudoscalar) is given by the scalar product between the photon's spin (a pseudovector) and the direction defined by the detector (a vector). Thus, the chiral reagent is substituted by the chiral observer (i.e. chiral setup). That is why we do not need to employ chiral properties of impinging electromagnetic fields. 

The role of the directionality of the detector in defining the handedness of the chiral setup highlights the crucial importance of having a vectorial response to the light-matter interaction, since a scalar response would be unable to exploit the directionality of the detector, and as a consequence also the handedness of the setup. Furthermore, as we show in Sec. \ref{sec:Symmetry}, such vectorial response automatically exhibits enantio-sensitivity and dichroism with respect to the \emph{external} vector defined by the detector. These properties indicate that in general the vectorial response results from the product of a molecular pseudoscalar and a field pseudovector. 
The field pseudovector determines the direction of observation of the dichroic and enantio-sensitive response and thus indicates (up to a sign) the corresponding detector arrangement required to measure such response (see Fig. \ref{fig:chiral_setup}).
The field pseudovector  is formed by non-collinear (and phase-delayed in the caes of a single frequency) components of the electric field. For example, in PECD, it results from the vector product between the $x$ and $y$ components of the circularly polarized field. Ultimately, the result of the measurement---the scalar (click)---is given by the projection of the vectorial response on the \emph{external} vector defined by detector, which yields the product of the molecular pseudoscalar and the handedness of the setup (see Sec. \ref{sec:vectorial_formulation}). The latter is  the projection (positive or negative) of the field pseudovector on the \emph{external} vector defined by the detector.

Note that the field pseudovector does not have to point in the direction of light propagation (as one might think from the above example). In Sec. \ref{sec:vectorial_formulation} we expose various opportunities 
offered by different field geometries, including 
arrangements of electric fields propagating non-collinearly.

In Sec. \ref{sec:vectorial_formulation} we illustrate this concept by deriving molecular pseudoscalars and field pseudovectors for four experiments detecting different  observables in different systems using different setups.  However, in all cases 
what enables chiral discrimination is the chiral observer defined by the combination of an achiral electromagnetic field and a directional detector.

\section{Symmetry in the electric-dipole approximation\label{sec:Symmetry}}

\begin{figure}
\noindent \begin{centering}
\includegraphics[scale=1.3]{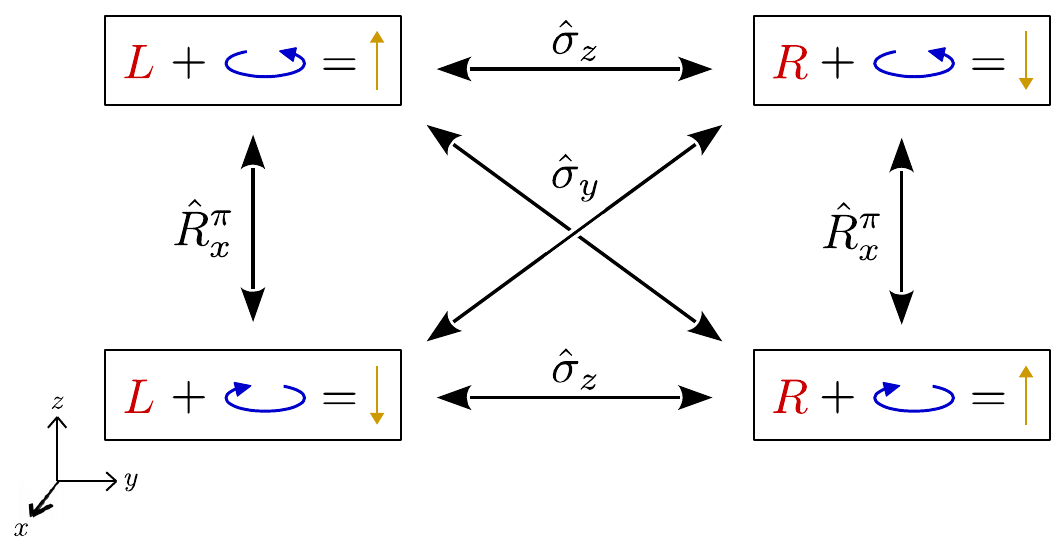}
\par\end{centering}

\caption{Symmetry properties of an isotropic ensemble of chiral molecules interacting
with circularly polarized light in the electric-dipole approximation. The box represents the ``enantiomer+field'' system. Inside the box: red letters $L$ and $R$ specify the
enantiomer, the curved blue arrow specifies the
direction of rotation of a field circularly polarized in the $xy$
plane, and the vertical golden arrow stands for a polar vector observable $\vec{v}=v_{z}\hat{z}$
displaying asymmetry with respect to the polarization $xy$ plane. A reflection
$\hat{\sigma}_{z}$ with respect to the $xy$ plane, leaves the field
invariant, but swaps the enantiomer and flips $\vec{v}$. A rotation
$\hat{R}_{\vec{a}}^{\pi}$ by $\pi$ radians around any axis $\vec{a}$
contained in the $xy$ plane leaves the enantiomer invariant because
the ensemble is isotropic, but swaps the polarization and flips $\vec{v}$. Note that a rotation $\hat{R}_{\vec{x}}^{\pi}$ ($\hat{R}_{\vec{y}}^{\pi}$) followed by a reflection $\hat{\sigma}_{z}$ is equivalent to a reflection $\hat{\sigma}_{y}$ ($\hat{\sigma}_{x}$) and leaves $\vec{v}$ invariant but swaps both the enantiomer and the polarization. \label{fig:Symmetry_circular}}

\end{figure}

\begin{figure}
\noindent \begin{centering}
\includegraphics[scale=1.3]{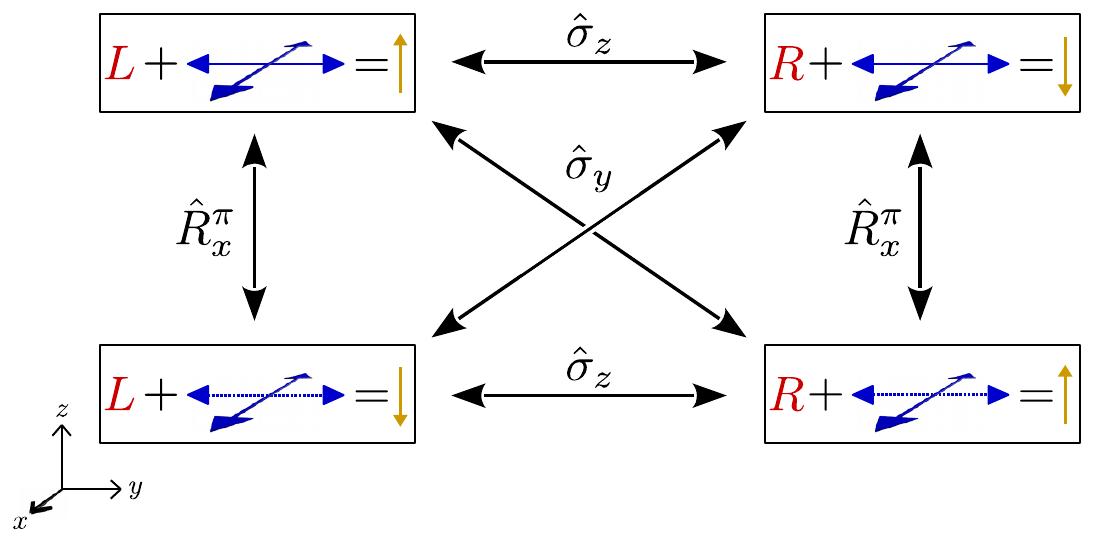}
\par\end{centering}

\caption{Same as Fig. \ref{fig:Symmetry_circular} but for two perpendicular
linearly polarized fields along $\hat{x}$ (double headed arrow in perspective)
and $\hat{y}$ (horizontal double headed arrow) of arbitrary frequencies
and intensities. In general $v_{x}\protect\neq0$ and $v_{y}\protect\neq0$
but only $v_{z}\hat{z}$ is shown (vertical arrow). A rotation $\hat{R}_{x}^{\pi}$
($\hat{R}_{y}^{\pi}$) leaves the enantiomer invariant but changes
the phase of the field along $\hat{y}$ ($\hat{x}$) by $\pi$.
$\hat{\sigma}_{x}$, $\hat{\sigma}_{y}$, $\hat{\sigma}_{z}$ describe transformations of the ``enantiomer+field'' system upon reflections with respect to the different axes of the lab frame.\label{fig:Symmetry_xy}}

\end{figure}

Let us begin with a simple symmetry consideration, which applies to all enantio-sensitive effects considered here. Consider first an isotropic
ensemble of a non-racemic mixture of chiral molecules, which interacts
with light circularly polarized in the $xy$ plane.
Irrespective of the specific chiral response we are looking
at, it may lead to an observable associated with some
polar vector $\vec{v}$. For example, in the case of PECD 
this polar vector is the net photoelectron current, 
while in PXCD it would be the coherent dipole induced
in the bound states of the neutral. 
\begin{figure}
\noindent \begin{centering}
\includegraphics[scale=1.3]{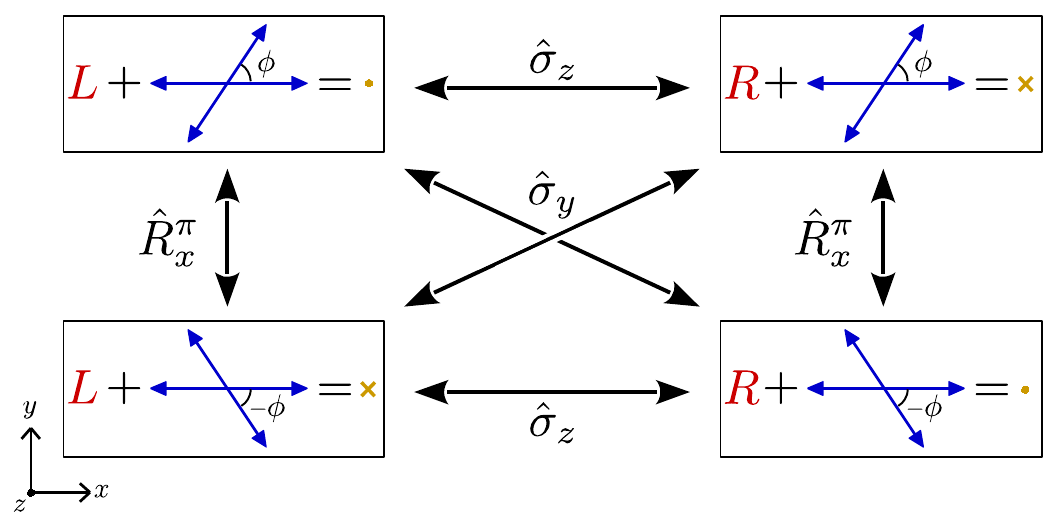}
\par\end{centering}

\caption{Same as Fig. \ref{fig:Symmetry_xy} but for an arbitrary angle between the two linearly polarized fields. Note that vectors pointing out of the page are indicated by a dot ($\hat{z}$ and $\vec{v}$ in the upper left and lower right configurations) and vectors pointing inside the page by an $\times$ ($\vec{v}$ in the upper right and lower left configurations).  
\label{fig:Symmetry_general}}

\end{figure}
The cylindrical symmetry of the ``ensemble+field'' system implies that $\vec{v}=v_{z}\hat{z}$ \footnote{Note that for few-cycle pulses, the cylindrical symmetry may be severely
compromised. However, for perturbative fields the first-order amplitudes
do not encode the duration of the pulse, that is, the response of
a few-cycle pulse can be emulated using monochromatic light of the
appropriate intensity, and therefore the cylindrical symmetry assumption
remains valid even for ultra-short pulses provided one only looks
at functions of the first-order amplitudes. }. Generalization to the 
case with no cylindrical symmetry is discussed below.
The ``enantiomer+field'' system and a chiral sensitive vectorial observable in the case of cylindrical symmetry are sketched
in the upper-left box of Fig. \ref{fig:Symmetry_circular}. It
applies, for example, to the field configuration in PECD and PXCD.
Our system transforms as indicated in Fig. \ref{fig:Symmetry_circular}
under reflections in the $xy$ plane and under rotations by $\pi$
radians around any axis contained in the $xy$ plane. These transformations
show the relationship between the different ``enantiomer+field'' configurations and the corresponding effect on the dichroic and enantio-sensitive observable $\vec{v}$. 

Figure \ref{fig:Symmetry_circular} shows that for an achiral 
ensemble, i.e. an ensemble of achiral molecules
or a racemic mixture of chiral molecules, the system ``ensemble+field''
is symmetric with respect to reflection $\hat{\sigma}_z$ in the $xy$ plane. Therefore, the vector $\vec{v}$ must vanish, yielding a photoelectron angular distribution symmetric with respect to the plane of polarization, otherwise two identical experiments would 
yield different results. However, for a non-racemic mixture
of chiral molecules, there is no symmetry enforcing $\vec{v}=0$.
Therefore, nothing prohibits the emergence of observables
which display asymmetry with respect to the plane of polarization,
 and the associated dichroism and enantio-sensitivity.
 The question is what these observables are, how strong can the signal be, and what determines its limits. We address these problems in the next section.

We also stress that the cylindrical symmetry is not essential for our reasoning. The argument can be extended to other geometries including linear fields or aligned molecules, provided one takes into account that $v_{x}$ and $v_{y}$ are not necessarily zero. Figure \ref{fig:Symmetry_xy}
shows a generalization of the case we have just considered. Now
the $x$ and $y$ components of the field have different
frequencies, intensities, and an arbitrary phase shift with respect
to each other. This field configuration is relevant, for example, for
the EMWS experiments carried out in Ref. \cite{patterson_sensitive_2013}.
The original experiment in Ref. \cite{patterson_enantiomer-specific_2013}
can also be analyzed similarly by replacing one of the two-headed
arrows in each ``enantiomer+field'' configuration in Fig. \ref{fig:Symmetry_xy}
by a single-headed arrow to account for the static field. The details 
of the analysis are discussed further in Sec. \ref{sub:AppendixStatic}, 
but the
conclusion remains the same: the emergence of a non-vanishing 
polar vector
characterizing the chiral response of the ``enantiomer+field'' system. Finally, the emergence of this vector for arbitrary orientations of linear fields is illustrated in Fig. \ref{fig:Symmetry_general}.

We can now support the introductory discussion of Sec. \ref{sec:chirality}
with several remarks concluding the symmetry analysis above:

First, 
from Figs. \ref{fig:Symmetry_circular}-\ref{fig:Symmetry_general} it is clear that 
$\vec{v}$ reflects the properties of the ``enantiomer+field'' system, and not those of the enantiomer or field separately.

Second, 
while it is well known that molecular chiral observables are characterized by pseudoscalars (scalar quantities that change sign upon the parity transformation), so far we have been discussing enantio-sensitive properties of a polar vector.
The appearance of a polar vector $\vec{v}$ is not accidental: its projections on the axes of the lab frame combine the information about the handedness of the chiral molecule and the handedness of the chiral setup. 

Third, the observation of enantio-sensitivity and dichroism in Figs. \ref{fig:Symmetry_circular}-\ref{fig:Symmetry_general} implicitly assumes a fixed $z$ direction against which we can compare the rotation direction of the light and the direction of the vector $\vec{v}$. Otherwise, there would be no way to, for example, distinguish right- and left-circularly-polarized light from each other, since we could rotate the $z$ axis by $\pi$ to change right- into left-circularly polarized light. Although a fixed $z$ direction is usually taken for granted, it remains physically meaningless until it is somehow related to the elements taking part in the experiment. In the methods we analyze here, such a $z$ direction is fixed by the detector (vertical arrow in Fig. \ref{fig:chiral_setup}), which is of course assumed to remain unchanged when either the enantiomer or the light polarization is changed. 

Therefore, the advent of electric-dipole-based techniques marks a shift of paradigm in chiral discrimination from using chiral reagents to using chiral observers, i.e. an experimental setup with well-defined handedness, even if the latter is not explicitly stated or recognized. 

In the next section 
we will show that in all cases the polar vector $\vec{v}$ is given by the product of (i) a molecular pseudoscalar and (ii) a field pseudovector specified by the configuration of the electric fields.
We will directly specify these two key quantities, forming the vectorial observables, for each of the electric-dipole-based techniques.

\section{Unified description  of chiral electric-dipole response}\label{sec:vectorial_formulation}

The chiral electric-dipole response manifests itself in vectorial observables, which have the following general form:
\begin{equation}
\vec{v}=\chi_m\vec{Z}_l,
\end{equation}
where $\chi_m$ is a molecular pseudoscalar defining the handedness of the molecule and $\vec{Z}_l$ is 
a light field pseudovector. Measuring $\vec{v}$ means projecting it on the \emph{external} vector $\vec{u}_d$ defined by the detector (vertical arrow in Fig. \ref{fig:chiral_setup}),
\begin{equation}
    \vec{v}\cdot\vec{u}_d = \chi_m \left(\vec{Z}_l \cdot{\vec{u}_d}\right).
    \label{eq:general_form}
\end{equation}
The projection of $\vec{Z}_l$ on $\vec{u}_d$ defines the handedness of the chiral setup, therefore, the result of the measurement is given by the product of the molecular handedness and the setup's handedness.
In this section we will derive $\chi_m$ and $\vec{Z}_l$ for four different electric-dipole based techniques of chiral discrimination. These techniques include PECD \cite{ritchie_theory_1976,cherepkov1982circular,powis_photoelectron_2000,bowering_asymmetry_2001}, EMWS \cite{patterson_enantiomer-specific_2013, patterson_sensitive_2013, Laane_2018},
PXCD \cite{beaulieu_PXCD_2016, *beaulieu_PXCD},
and PXECD \cite{beaulieu_PXCD_2016, *beaulieu_PXCD}. 

\subsection{Photoelectron circular dichroism}
We begin with what is 
perhaps the most prominent 
electric-dipole-based technique, PECD. This technique was first proposed in 1976 \cite{ritchie_theory_1976} and  then rediscovered in 1982 \cite{cherepkov1982circular}. The first quantitative calculations of the effect \cite{powis_photoelectron_2000} yielded staggering results: the expected  effect was at the level of some few percent to maybe even some ten percent of the total photoionization signal. The first experiment appeared just a year later \cite{bowering_asymmetry_2001}. The technique was dramatically advanced in Refs. \cite{powis00,powis2008giant,nahon_determination_2006,nahon_valence_2015,Nahon2016_det}
from a theoretical concept to an extra-sensitive experimental technique.
With the advances in table-top 
laser-based implementations \cite{fanood2015enantiomer,rafiee2016wavelength}
including multi-photon \cite{lux_circular_2012,lehmann_imaging_2013,lux_photoelectron_2015} and strong-field regimes \cite{beaulieu_universality_2016},
PECD has proven  very interesting from both fundamental and applied perspectives. 
In PECD, the photoionization of an isotropic and non-racemic ensemble
of chiral molecules by circularly polarized light leads to an asymmetry in the photoelectron angular distribution (PAD) with respect to the polarization
plane, the so-called  forward-backward
asymmetry (FBA). This asymmetry is usually described by decomposing the angle-resolved photoionization
probability $W(\vec{k}^{\mathrm{L}})$ in Legendre polynomials,
\begin{equation}
W(\vec{k}^{\mathrm{L}})=\sum_{l=0}^{2}b_{l}\left(k\right)P_{l}\left(\cos\theta_{k}^{\mathrm{L}}\right),\label{eq:W(k^L)}
\end{equation}
where it corresponds to a non-zero $b_{1}$ coefficient. In Eq. \eqref{eq:W(k^L)},
$W(\vec{k}^{\mathrm{L}})$ is the probability of obtaining
a photoelectron with momentum $\vec{k}^{\mathrm{L}}$, $\mathrm{L}$
indicates that the vector is in the lab frame, $\theta_{k}^{\mathrm{L}}$
is the polar angle of $\vec{k}^{\mathrm{L}}$, $k\equiv\vert\vec{k}^\mathrm{L}\vert$, $P_{l}$ is the Legendre
polynomial of degree $l$, and we assume that the polarization plane
coincides with the $x^{\mathrm{L}}y^{\mathrm{L}}$ plane. The $b_{1}$
coefficient is directly related to the net photoelectron current induced by ionization 
\begin{equation}
\vec{j}^{\mathrm{L}}\left(k\right)=\int\mathrm{d}\Omega_{k}^{\mathrm{L}}\vec{j}^{\mathrm{L}}(\vec{k}^{\mathrm{L}}),\label{eq:current}
\end{equation}
where $\vec{j}^{\mathrm{L}}(\vec{k}^{\mathrm{L}})=W(\vec{k}^{\mathrm{L}})\vec{k}^{\mathrm{L}}$ is the photolectron current in the direction specified by the photoelectron direction $\vec{k}^{\mathrm{L}}$ in the lab frame,  and $\int\mathrm{d}\Omega_{k}^{\mathrm{L}}\equiv\int_{0}^{\pi}\mathrm{d}\theta_{k}^{\mathrm{L}}\int_{0}^{2\pi}\mathrm{d}\varphi_{k}^{\mathrm{L}}\sin\theta_{k}^{\mathrm{L}}$ is the integral over all photoelectron directions. From the orthogonality of the Legendre polynomials we obtain 
\begin{eqnarray}
\vec{j}^{\mathrm{L}}\left(k\right)&=&\int\mathrm{d}\Omega_{k}^{\mathrm{L}}\vec{j}^{\mathrm{L}}(\vec{k}^{\mathrm{L}}),\nonumber\\
&=&\sum_{l=0}^{2}b_{l}(k)\int\mathrm{d}\Omega_{k}^{\mathrm{L}}P_{l}\left(\cos\theta_{k}^{\mathrm{L}}\right)\vec{k}^{\mathrm{L}},\nonumber\\
&=&k\sum_{l=0}^{2}b_{l}(k)\int\mathrm{d}\Omega_{k}^{\mathrm{L}}P_{l}\left(\cos\theta_{k}^{\mathrm{L}}\right)P_{1}\left(\cos\theta_{k}^{\mathrm{L}}\right)\hat{z}^{\mathrm{L}},\nonumber\\
&=&\frac{4\pi}{3}kb_{1}(k)\hat{z}^{\mathrm{L}},\label{eq:j^L_and_b1}
\end{eqnarray}

The current in Eqs. \eqref{eq:current} and \eqref{eq:j^L_and_b1} is the vectorial observable of our interest. 
The task is to find it or, equivalently, $b_1(k)$. 
The corresponding calculations of the photoelectron angular distributions traditionally 
rely on the formalism of angular momentum algebra,
both for one-photon and few-photon PECD \cite{ritchie_theory_1976, cherepkov1982circular, powis_photoelectron_2000, Koch2017}. We have found that it is very instructive to depart from this traditional formalism, which uses language specific for photoionization.
Instead, we use an alternative, vectorial formulation, pioneered in
works of Manakov \cite{manakov1996} and applied  to aligned
chiral systems \cite{agre2006}. The vectorial formalism was also used  to describe two-photon absorption CD  \cite{tinoco1975}. 
Conveniently, it provides a common language for all electric-dipole-based techniques, irrespective of their ``field of origin'' or observable, be it photoionization or microwave physics.

We define the incident circularly polarized field in the lab reference frame as 
\begin{equation}
\vec{E}(t)=\mathcal{E}(t)\hat{e}_{\sigma}^{\mathrm{L}}+\mathrm{c.c.}
\label{eq:field}
\end{equation}
where $\hat{e}_{\sigma}^{\mathrm{L}}=\left(\hat{x}^{\mathrm{L}}+\mathrm{i}\sigma\hat{y}^{\mathrm{L}}\right)/\sqrt{2}$
is the light polarization vector, $\sigma=\pm 1$ defines the rotation
direction of the field, and $\mathcal{E}(t)$ is the time-dependent
amplitude. 
The photoelectron
current density for a given photoelectron momentum $\vec{k}^{\mathrm{M}}$
in the molecular frame is (up to the negative electron charge)
\begin{equation}
\vec{j}_{\vec{k}^{\mathrm{M}}}^{\mathrm{M}}=\left|a_{\vec{k}^{\mathrm{M}}}\right|^{2}\vec{k}^{\mathrm{M}}.\label{eq:j_mol}
\end{equation}
Here $\vec{a}_{\vec{k}^{\mathrm{M}}}$ is the ionization amplitude of the transition to the continuum state $\vert\vec{k}^{\mathrm{M}}\rangle$
from the ground state $\vert0\rangle$ in the circularly polarized
field Eq. \eqref{eq:field}. Its standard first-order perturbation theory
expression is
\begin{equation}
a_{\vec{k}^{\mathrm{M}}}
=\i\tilde{\mathcal{E}}\braoket{\vec{k}^{\mathrm{M}}}{\vec{d}^{\mathrm{L}}\t\hat{e}_{\sigma}^{\mathrm{L}}}0
=\frac{\i\tilde{\mathcal{E}}}{\sqrt{2}}\left(\vec{D}^{\mathrm{L}}\t\hat{x}^{\mathrm{L}}+\sigma\i\vec{D}^{\mathrm{L}}\t\hat{y}^{\mathrm{L}}\right),
\label{eq:amplitude}
\end{equation}
where $\tilde{\mathcal{E}}$ is the Fourier transform of $\mathcal{E}$
at the transition frequency, $\vec{d}$ is the dipole operator, and $\vec{D}^{\mathrm{L}}$
is the corresponding transition dipole matrix element
in the lab frame.

Our next step is to identify the  molecule-specific enantio-sensitive structure in Eqs. \eqref{eq:j_mol} and \eqref{eq:amplitude}. That is, we will be looking for molecule-specific pseudoscalars; quantities that change sign upon parity inversion. Pseudoscalars may arise as a product of a vector and pseudovector. An example of such pseudoscalar is the helicity $\eta$ of circularly polarized light which is non-zero only beyond the electric-dipole approximation (see Appendix \ref{sec:AppendixCD}). Molecular pseudoscalars also arise from triple products formed by three molecular polar vectors. We shall now look for such quantities. 

Let us look at the cross term arising in  $|a_{\vec{k}^{\mathrm{M}}}|^2$,
\begin{equation}
\i\sigma
\left[(\vec{D}^{\mathrm{L}*}\t\hat{x}^{\mathrm{L}})
(\vec{D}^{\mathrm{L}}\t\hat{y}^{\mathrm{L}})-
(\vec{D}^{\mathrm{L}}\t\hat{x}^{\mathrm{L}})
(\vec{D}^{\mathrm{L*}}\t\hat{y}^{\mathrm{L}})
\right].
\label{eq:crossterm}
\end{equation}
We now use the vector identity $(\vec{a}\t\vec{c})(\vec{b}\t\vec{d})-(\vec{a}\t\vec{d})(\vec{b}\t\vec{c})=(\vec{a}\x\vec{b})\t(\vec{c}\x\vec{d})$
and the fact that $\hat{x}^{\mathrm{L}}\x \hat{y}^{\mathrm{L}}=\hat{z}^{\mathrm{L}}$
to write the interference term as a triple product, 
\begin{equation}
\left|a_{\vec{k}_{\mathrm{M}}}\right|^{2}=\frac{\left|\tilde{\mathcal{E}}\right|^{2}}{2}\bigg\{\left|\vec{D}^{\mathrm{L}}\t\hat{x}^{\mathrm{L}}\right|^{2}+\left|\vec{D}^{\mathrm{L}}\t\hat{y}^{\mathrm{L}}\right|^{2}+\i\sigma\left(\vec{D}^{\mathrm{L}*}\x\vec{D}^{\mathrm{L}}\right)\t\hat{z}^{\mathrm{L}}\bigg\}.\label{eq:a_k_tripleproduct}
\end{equation}
Note that $\i\left(\vec{D}^{*}\x\vec{D}\right)=2\Im\{\vec{D}\}\x\Re\{\vec{D}\}$ is a real
vector, where $\Re\{\vec{D}\}$ and $\Im\{\vec{D}\}$ are the real and imaginary parts of $\vec{D}$.

The last term in Eq. (\ref{eq:a_k_tripleproduct}) is a triple product, but it is not the one  we were looking for. Indeed, instead of a polar vector, $\sigma\hat{z}^\mathrm{L}$ is a pseudovector that characterizes the rotation direction of the field, i.e. the photon's spin (see Appendix \ref{sec:AppendixCD}), and moreover, the triple product 
includes two vectors characterizing the molecule and one vector characterizing the ``observer'' (or the lab frame), as opposed to three vectors characterizing the molecule in the molecular frame.

To relate the above  expression to 
the transition dipoles in the molecular, rather than the lab frame, one can use the rotation matrix 
$S(\varrho)$. It transforms the vectors
from the molecular to the lab frame
via a rotation through the Euler angles $\varrho\equiv\left(\alpha\beta\gamma\right)$:  $\vec{D}^{\mathrm{L}}=S\vec{D}^{\mathrm{M}}\equiv S\langle\vec{k}^{\mathrm{M}}\vert\vec{d}^{\mathrm{M}}\vert0\rangle$.

Using Eq. (\ref{eq:a_k_tripleproduct}), we can also write the current in the lab frame, corresponding to the photoelectron
momentum $\vec{k}^{\mathrm{M}}$ in the molecular frame
\begin{equation}
\vec{j}_{\vec{k}^{\mathrm{M}}}^{\mathrm{L}}=S\vec{j}_{\vec{k}^{\mathrm{M}}}^{\mathrm{M}}=\frac{\left|\tilde{\mathcal{E}}\right|^{2}}{2}\left[\left|S\vec{D}^{\mathrm{M}}\t\hat{x}^{\mathrm{L}}\right|^{2}+\left|S\vec{D}^{\mathrm{M}}\t\hat{y}^{\mathrm{L}}\right|^{2}+\sigma\i S\left(\vec{D}^{\mathrm{M}*}\x\vec{D}^{\mathrm{M}}\right)\t\hat{z}^{\mathrm{L}}\right]S\vec{k}^{\mathrm{M}}.
\label{eq:j_lab}
\end{equation}
Note that $\vec{D}^{\mathrm{L}*}\x\vec{D}^{\mathrm{L}}=S(\vec{D}^{\mathrm{M}*}\x\vec{D}^{\mathrm{M}})$.
This current is not a usual observable. Measuring it would require
a coincidence-type setup, where one would  detect the lab-frame electron momentum together with the orientation  of the molecular frame in the lab frame.
We are interested in the standard observable -- the 
net photoelectron current in the lab frame. Therefore, we need to integrate over all directions of the photoelectron momentum
and over all molecular orientations:

\begin{eqnarray}
\vec{j}^{\mathrm{L}}\left(k\right) & = & \int\mathrm{d}\varrho\int\mathrm{d}\Omega_{k}^{\mathrm{M}}\vec{j}_{\vec{k}^{\mathrm{M}}}^{\mathrm{L}},
\label{eq:j_lab_total}
\end{eqnarray}
where $\int\mathrm{d}\Omega_{k}^{\mathrm{M}}\equiv\int_{0}^{\pi}\mathrm{d}\theta_{k}^{\mathrm{M}}\int_{0}^{2\pi}\mathrm{d}\varphi_{k}^{\mathrm{M}}\sin\theta_{k}^{\mathrm{M}}$.

Within the standard approach, one performs the integration over all molecular orientations keeping the photoelectron momentum $\vec{k}$ fixed in the lab frame. This yields the standard lab-frame photoelectron angular distributions, from which the $b_1$ coefficient, which is proportional to the net photoelectron current [see Eq. \ref{eq:j^L_and_b1}], is extracted. Here, since we are not interested in the full angular distribution of photoelectrons, we can keep the photoelectron momentum $\vec{k}$ fixed in the molecular frame. This simplifies the orientation averaging procedure considerably because 
in this case the transition matrix element vector $\vec{D}^{\mathrm{M}}(\vec{k}^{\mathrm{M}})$ does not have an argument that depends on the molecular orientation $\varrho$, and can therefore be trivially rotated as $S(\varrho)\vec{D}^{\mathrm{M}}(\vec{k}^{\mathrm{M}})$. In the other case, when $\vec{k}$ is fixed in the lab frame, the corresponding rotation reads as $S(\varrho)\vec{D}^{\mathrm{M}}(S(\varrho)^{-1}\vec{k}^{\mathrm{L}})$ and the orientation averaging step requires knowing how $\vec{D}$ changes as a function of $\vec{k}$, which is usually tackled with a partial wave expansion of the continuum wave function. We do not have such complication here and we can simply use the vector identitiy Eq. \eqref{eq:(a.b)v} derived 
in Appendix \ref{sec:AppendixOrientationAveraging} to 
obtain

\begin{eqnarray}
\vec{j}^{\mathrm{L}}\left(k\right) = \left\{ \frac{1}{6}\int\mathrm{d}\Omega_{k}^{\mathrm{M}}\left[\i\left(\vec{D}^{\mathrm{M}*}\x\vec{D}^{\mathrm{M}}\right)\t\vec{k}^{\mathrm{M}}\right]\right\} \left\{ \sigma\left|\tilde{\mathcal{E}}\right|^{2}\hat{z}^{\mathrm{L}}\right\} . \label{eq:j_lab_total_factored}
\end{eqnarray}

The equivalence between expression \eqref{eq:j_lab_total_factored}
and the original expression derived by Ritchie in \cite{ritchie_theory_1976} is demonstrated 
in Appendix \ref{sub:AppendixRitchie}. 

Expression \eqref{eq:j_lab_total_factored} is  physically transparent. In particular, it
shows that the strength of the chiral signal depends on
the mutual orientation of the three vectors forming the 
triple product of vectors defined in the molecular frame.   

Let us analyze expression \eqref{eq:j_lab_total_factored}: 

First,  we see that 
only the interference term in the current 
[see Eq. (\ref{eq:j_lab})]
yields a non-vanishing contribution to the net current after orientation
averaging. This stresses the importance of the coherence between the two contributions to the ionization amplitude, triggered by the two components of the ionizing
field. 

Second, we see that the orientation averaging has modified the expression for the vector triple product: it no longer involves
any lab-frame quantities, such as $\sigma \hat z^{\rm L}$. Its
place is now taken by the molecular frame photoelectron
momentum $\vec{k}^{\mathrm{M}}$, and the molecular term is now a rotationally invariant quantity.

Third, Eq. \eqref{eq:j_lab_total_factored} shows that the net photoelectron
current (per molecule) in the lab frame can be factored into a pseudovector field term expressed in the lab frame and a pseudoscalar molecular term expressed in the molecular frame.
The pseudovector field term contains the intensity of the field at the
transition frequency, and the rotation direction
of the circularly polarized field $\sigma \hat{z}^\mathrm{L}$. 
The molecular term is an integral over all states on the 
photoelectron energy shell $k^{2}/2$, where, after taking into account all molecular
orientations, each state contributes by an amount proportional to
the scalar triple product between $\vec{D}^{\mathrm{M}}(\vec{k}^{\mathrm{M}})$,
$\vec{D}^{\mathrm{M}*}(\vec{k}^{\mathrm{M}})$, and $\vec{k}^{\mathrm{M}}$,
or equivalently between $\Re\{\vec{D}^{\mathrm{M}}(\vec{k}^{\mathrm{M}})\}$,
$\Im\{\vec{D}^{\mathrm{M}}(\vec{k}^{\mathrm{M}})\}$, and $\vec{k}^{\mathrm{M}}$. 

From the field term we can see that $\vec{j}^{\mathrm{L}}(k)$ is
directed along $\hat{z}^{\mathrm{L}}$ and takes opposite values for
opposite circular polarizations and a given enantiomer. 
On the other
hand, from the relationship between the photoionization dipoles of opposite
enantiomers derived in Appendix \ref{sub:AppendixDipole}, $\vec{D}_{\mathrm{left}}^{\mathrm{M}}(\vec{k}^{\mathrm{M}})=-\vec{D}_{\mathrm{right}}^{\mathrm{M}}(-\vec{k}^{\mathrm{M}})$,
it is simple to see that the molecular term is a pseudoscalar, i.e. it changes
sign under a parity inversion, and therefore $\vec{j}^{\mathrm{L}}(k)$
takes opposite values for the 
opposite enantiomers and a given circular
polarization [see Eqs. \eqref{eq:D_R(k)=-D_L(-k)} and \eqref{eq:chi_R=-chi_L} in Appendix \ref{sub:AppendixDipole}]. All these conclusions are in agreement with 
the symmetry analysis described in Sec.
\ref{sec:Symmetry}, with $\vec{j}^{\mathrm{L}}(k)$ 
playing the role of the generic dichroic and enantio-sensitive vector $\vec{v}$. 

The triple product in the molecular term vanishes if the vectors are
coplanar, which is for example the case for the plane wave continuum, where one can use the velocity gauge to show
that $\vec{D}^{\mathrm{M}}$ is parallel to $\vec{k}^{\mathrm{M}}$. This conclusion 
corresponds to the well known fact that $\vert\vec{j}^{\mathrm{L}}\left(k\right)\vert/k\propto\vert b_{1}\vert$
has an overall tendency to decrease as the photoelectron energy increases
and the continuum resembles more and more a plane wave.
One can also show that $\vec{j}^{\mathrm{L}}(k)$  vanishes in case of  a spherically symmetric continuum in agreement with earlier studies \cite{cherepkov1982circular}. The same conclusion holds for the strong-field PECD \cite{lein_2014}.  

Our derivation and the result provide us with an important
insight. The chiral signal stems from the interference between the two
non-collinear dipole transitions. If we consider a single final state,
such interference leading to a vector product of two transition dipoles
would only be possible for a scattering state where the complex transition
dipole allows for two non-collinear components: one of them is given
by the real part of the transition dipole and the other by its imaginary part. 

The generalization of Eq. \eqref{eq:j_lab_total_factored} to arbitrary polarizations of the field is straightforward. We just need to separate the Fourier transform of the field into its real and imaginary parts, and keep in mind that for any complex vector $\vec{u}=\vec{u}_{r}+\i\vec{u}_{i}$ we have that $\vec{u}^{*}\x\vec{u}=-2\i\vec{u}_{i}\x\vec{u}_{r}$. Then we obtain

\begin{equation}
\vec{j}^{\mathrm{L}}\left(k\right)=\left\{ \frac{1}{6}\int\mathrm{d}\Omega_{k}^{\mathrm{M}}\left[\left(\vec{D}^{\mathrm{M}*}\x\vec{D}^{\mathrm{M}}\right)\t\vec{k}^{\mathrm{M}}\right]\right\} \left\{ \tilde{\vec{\mathcal{E}}}^{\mathrm{L}*}\x\tilde{\vec{\mathcal{E}}}^{\mathrm{L}}\right\}, \label{eq:j_lab_total_factored_general}
\end{equation}

 which reduces to Eq. \eqref{eq:j_lab_total_factored} for the case of circularly polarized light. Eq. \eqref{eq:j_lab_total_factored_general} shows that for an arbitrary field configuration the chiral response in PECD is not necessarily along the light propagation direction. 

\subsection{Photo-excitation circular dichroism in electronic or  vibronic states}\label{sub:Bound-bound-transitions}

Let us now consider chiral response in bound excited states. In this case, and for a single excited state, the excitation
dipole is real. Therefore, 
$\vec{D}^{\mathrm{M}}(\vec{k}^{\mathrm{M}})$ and
$\vec{D}^{\mathrm{M}*}(\vec{k}^{\mathrm{M}})$ are parallel, yielding zero enantio-sensitive dipole signal. 

On the other hand, if we were to coherently excite two states with non collinear transition dipoles, we would have a non-zero cross product. Then we could obtain a dichroic and enantio-sensitive signal as long as we find a vectorial signal that involves the interference between the two excitations. Unlike in the previous case where this vectorial signal was provided by the photoelectron current, in this case, it is provided by the dynamics of the induced polarization.

The goal of our analysis is to uncover the intimate connection between the PXCD effect discovered in \cite{beaulieu_PXCD_2016, *beaulieu_PXCD}
for electronic and vibronic states and the 
EMWS
discovered in \cite{patterson_enantiomer-specific_2013}
for the rotational states. The physics in these two cases is quite different, as the former involves internal and the 
latter external degrees of freedom, leading
to subtle but important details in the  
mathematical treatment.

Consider the case of two electronic or vibronic states, which 
can be coherently excited by an ultrashort pulse from the ground
state. 
As before, we will consider a randomly oriented ensemble. After interaction
with a field of arbitrary frequency, polarization, and intensity,
the first-order 
amplitudes of the excited states are given by

\begin{equation}
a_{j}\left(t\right)=\i\left[\vec{d}_{j,0}^{\mathrm{L}}\t\tilde{\vec{\mathcal{E}}}^{\mathrm{L}}\left(\omega_{j0}\right)\right]\e^{-\i\omega_{j}t},\qquad j=1,2.
\label{eq:BoundAmplitudes}
\end{equation}

where $\vec{d}_{j,0}^{\mathrm{L}}$ is now the real-valued transition dipole between
the ground and $j$-th excited state and $\tilde{\vec{\mathcal{E}}}^{\mathrm{L}}$
is the Fourier transform of the field at the
corresponding transition frequency. For an ultrashort pulse
with the bandwidth covering both excited states, 
the expectation value of the dipole
will contain an interference term of the form

\begin{eqnarray}
\langle\vec{d}^{\mathrm{L}}\rangle_{\chi} & \equiv & a_{1}^{*}a_{2}\vec{d}_{1,2}^{\mathrm{L}}+\mathrm{c.c.}\nonumber \\
 & = & \left[\vec{d}_{0,1}^{\mathrm{L}}\t\tilde{\vec{\mathcal{E}}}^{\mathrm{L}*}\left(\omega_{1,0}\right)\right]\left[\vec{d}_{2,0}^{\mathrm{L}}\t\tilde{\vec{\mathcal{E}}}^{\mathrm{L}}\left(\omega_{2,0}\right)\right]\vec{d}_{1,2}^{\mathrm{L}}\e^{-\i\omega_{2,1}t}+\mathrm{c.c.}
 \label{eq:metadipole}
\end{eqnarray}
which we have denoted by $\langle\vec{d}^{\mathrm{L}}\rangle_{\chi}$
to indicate that it is the chiral part of the induced polarization.

In contrast to Eq. \eqref{eq:a_k_tripleproduct} and PECD, 
the fact that the Fourier transform of the field is evaluated at two different transition frequencies in the above expression does not allow us
to  easily  use the vector identity 
$(\vec{a}\t\vec{c})(\vec{b}\t\vec{d})-(\vec{a}\t\vec{d})(\vec{b}\t\vec{c})=(\vec{a}\x\vec{b})\t(\vec{c}\x\vec{d})$ and
directly identify a triple product. The emergence of the
triple-product as an enantio-sensitive measure is somewhat subtle: it only appears after  averaging over {\it all } molecular orientations, for a randomly oriented molecular ensemble. 
With the help of Eq. \eqref{eq:=00005B(axb).c=00005D(uxv)} derived in Appendix \ref{sec:AppendixOrientationAveraging}, one finds that
\begin{equation}
\int\mathrm{d}\varrho\left\langle \vec{d}^{\mathrm{L}}\right\rangle _{\chi}=\frac{1}{6}\left[\left(\vec{d}_{0,1}^{\mathrm{M}}\x\vec{d}_{2,0}^{\mathrm{M}}\right)\t\vec{d}_{1,2}^{\mathrm{M}}\right]\left[\tilde{\vec{\mathcal{E}}}^{\mathrm{L}*}\left(\omega_{1,0}\right)\x\tilde{\vec{\mathcal{E}}}^{\mathrm{L}}\left(\omega_{2,0}\right)\right]\e^{-\i\omega_{2,1}t}+\mathrm{c.c.}\label{eq:PXCD_general}
\end{equation}

The essential features of this expression are similar to those of PECD. The expression again factorizes into a molecular part, a pseudoscalar given by the triple product of molecule-specific transition dipoles, and a field part, a pseudovector given by the vector product of the incident fields. The 
induced dipole oscillates at the frequency $\omega_{2,1}$ in the
direction determined by the cross product between the Fourier transforms
of the exciting fields, at the corresponding transition frequencies.
The triple product of the transition dipoles is taken in the molecular frame and forms the pseudoscalar that changes sign for opposite enantiomers (see Appendix \ref{sub:AppendixDipole}). 
This means that the phase of the oscillations will be determined by the product of the signs resulting from the molecular and field terms. For a fixed polarization and opposite enantiomers, or for a fixed enantiomer and opposite polarizations (see Figs. \ref{fig:Symmetry_circular}-\ref{fig:Symmetry_general}), the  phase will change by $\pi$. That is, the enantio-sensitive and dichroic character of the vectorial observable, in this case the polarization, is encoded in the phase of its oscillations.

In the particular case of a circularly polarized field [see 
Eq.\eqref{eq:field}], we have $\tilde{\vec{\mathcal{E}}}^{\mathrm{L}}\left(\omega\right)=\tilde{\mathcal{E}}\left(\omega\right)\left(\hat{x}^{\mathrm{L}}+\sigma\i\hat{y}^{\mathrm{L}}\right)/\sqrt{2}$
and therefore

\begin{equation}
\int\mathrm{d}\varrho\left\langle \vec{d}^{\mathrm{L}}\right\rangle _{\chi}=\frac{\i\sigma}{6}\left[\left(\vec{d}_{0,1}^{\mathrm{M}}\x\vec{d}_{2,0}^{\mathrm{M}}\right)\t\vec{d}_{1,2}^{\mathrm{M}}\right]\tilde{\mathcal{E}}^{*}\left(\omega_{1,0}\right)\tilde{\mathcal{E}}\left(\omega_{2,0}\right)\hat{z}^{\mathrm{L}}\e^{-\i\omega_{2,1}t}+\mathrm{c.c.},
\label{eq:PXCD}
\end{equation}

which is the PXCD effect discovered in \cite{beaulieu_PXCD_2016, *beaulieu_PXCD}.

Equation \eqref{eq:PXCD_general} is the generalization of the PXCD effect
to the case of an arbitrary field. It shows that one can obtain the same effect
by either using a single broadband elliptically polarized pulse or,
for example, by using a sequence of two spectrally narrow (and phase locked) linearly polarized pulses with
orthogonal polarizations. If more than two levels are coherently excited, then Eq. \eqref{eq:PXCD} should include the sum over all states.

Importantly,  the vectorial quantity associated with the chiral response does not have to be collinear with the direction of light propagation, as it happens in the case of a circularly polarized field. It illustrates once again, that the light propagation direction, fundamental for characterizing the chirality of a photon, does not play any role in electric-dipole-based techniques. These techniques do not use the chirality of the photon, but use the polarization vectors of the light to define the lab setup.

An important feature that distinguishes the ``light-observer'' from the ``light-reagent'' is the presence of chiral sensitive absorption. 
Of course, PECD is associated with light absorption, but this absorption is not chiral sensitive, e.g. it is neither enantio-sensitive nor dichroic \cite{beaulieu_PXCD_2016, *beaulieu_PXCD}.

Note that the earlier results for the quadratic susceptibility in isotropic chiral media 
can also be presented in the vectorial form, originally derived by Giordmaine \cite{giordmaine_1965},

\begin{equation}
\vec{P}(\omega_3=\omega_1-\omega_2) =\chi^{(2)}[\vec{E}_1(\omega_1)\times \vec{E}^{*}_{2}(\omega_2)],
\end{equation}

where the vectors 
$\vec{P}$, $\vec{E}_1$, and $\vec{E}_2$, 
are the Fourier components of induced polarizations and incident fields at the respective frequencies,  $\chi^{(2)}$ is the molecular pseudoscalar described by the triple product of transition dipoles and a combination of resonance denominators typical for second order instantaneous response and derived in 
\cite{fischer_three-wave_2000,fischer2001isotropic} in the context of tree-wave mixing in isotropic chiral media within the electric-dipole approximation. 

Finally, the  expression for PXCD also allows one to gauge the strength of
the chiral response. It maximizes 
when the three transition dipoles are orthogonal to each other. 
In this case, the coherent enantio-sensitive dipole along the
lab $\hat z^{\textrm L}$ axis, normalized to the excitation amplitudes, reaches $d_{1,2}^{\textrm M}/3$. 
Thus, for orthogonal excitation dipoles, the molecule can convert all of its (ensemble-averaged) initial excitation in the polarization plane of the circularly polarized pump  into enantio-sensitive motion orthogonal 
to this plane, making  a highly efficient helix.

\subsection{Chiral response upon rotational excitation: enantio-sensitive microwave spectroscopy} \label{sec:microwavePXCD}

In this section we will use our vectorial formulation to consider two enantio-sensitive schemes in the microwave regime suggested by Patterson et al. \cite{patterson_enantiomer-specific_2013, patterson_sensitive_2013}, and described theoretically in detail by Lehman \cite{Laane_2018}. 

Consider first coherent excitation of rotational states
and the enantio-sensitive signal discovered  by Patterson et al. in \cite{patterson_sensitive_2013}.
The corresponding rotational wavefunctions are the eigenstates of the asymmetric
rigid rotor \cite{kroto_molecular_1992}, and are 
themselves functions of the Euler angles. 
We no longer deal with  {\it 
a posteriori} averaging over this degree of freedom. The transition
dipoles themselves are already the
integrals over the Euler angles $\varrho\equiv\left(\alpha\beta\gamma\right)$,
\begin{eqnarray}
\vec{d}_{i,j}^{\mathrm{L}} & = & \bra{J_{i}\tau_{i}M_{i}}\vec{d}^{\mathrm{L}}\ket{J_{j}\tau_{j}M_{j}}\nonumber \\
 & = & \left[\int\mathrm{d}\varrho\,\psi_{J_{i}\tau_{i}M_{i}}^{*}\left(\varrho\right)S\left(\varrho\right)\psi_{J_{j}\tau_{j}M_{j}}\left(\varrho\right)\right]\vec{d}^{\mathrm{M}},
\end{eqnarray}
where $\vec{d}^{\mathrm{M}}$ is the permanent dipole moment of the electronic ground state  in the
molecular frame, $i,j=1,2$. The state $\ket{J \tau M }$ is an eigenfunction of the total angular momentum operator $\hat{J}^2$ and its $z$-component $\hat{J_z}$ with eigenvalues $J(J+1)$ and $M$, respectively, and $\tau$ is associated with all other quantum numbers pertinent for this state. 
These transition dipoles are now used for the 
excitation amplitudes, which are still given by 
the general expression Eq.  (\ref{eq:BoundAmplitudes}) and the induced dipole Eq. (\ref{eq:metadipole}).
Each of the dipoles entering Eq. (\ref{eq:metadipole}) is associated with a distribution of possible $M_i,M_j,M_k$. This distribution depends on the preparation of the system.  

The orientation averaging over the Euler angles is now 
replaced by summing over the distribution of 
all possible initial and final $M$'s

\begin{eqnarray}
\sum_{M_{0},M_{1},M_{2}}\left\langle \vec{d}^{\mathrm{L}}\right\rangle _{\chi} 
 & = & \sum_{M_{0},M_{1},M_{2}}\left[\vec{d}_{0,1}^{\mathrm{L}}\t\tilde{\vec{\mathcal{E}}}^{\mathrm{L}*}\left(\omega_{1,0}\right)\right]\left[\vec{d}_{2,0}^{\mathrm{L}}\t\tilde{\vec{\mathcal{E}}}^{\mathrm{L}}\left(\omega_{2,0}\right)\right]\vec{d}_{1,2}^{\mathrm{L}}\e^{-\i\omega_{2,1}t}+\mathrm{c.c.}
 \label{eq:metadipole1}
\end{eqnarray}

When all possible
initial and final $M$'s are equally represented, as is the case for an isotropic sample,
the averaging is performed with the help of Eq.
(\ref{eq:(A.u)(B.v)C =00003D =00005B(AxB).C=00005D=00005Buxv=00005D/6}) which is derived the Appendix \eqref{sub:Appendix_QuantumOrientationAveraging}
and yields 
\begin{equation}
\sum_{M_{0},M_{1},M_{2}}\left\langle \vec{d}^{\mathrm{L}}\right\rangle _{\chi}=\sum_{M_{0},M_{1},M_{2}}\frac{1}{6}\left[\left(\vec{d}_{0,1}^{\mathrm{M}}\x\vec{d}_{2,0}^{\mathrm{M}}\right)\t\vec{d}_{1,2}^{\mathrm{M}}\right]\left[\tilde{\vec{\mathcal{E}}}^{\mathrm{L}*}\left(\omega_{1,0}\right)\x\tilde{\vec{\mathcal{E}}}^{\mathrm{L}}\left(\omega_{2,0}\right)\right]\e^{-\i\omega_{2,1}t}+\mathrm{c.c.}\label{eq:d_chi_rotational_states}
\end{equation}
The main result here is the factorization of induced polarization into the molecular-specific pseudoscalar $\sum_{M_{0},M_{1},M_{2}}\frac{1}{6}\left[\left(\vec{d}_{0,1}^{\mathrm{M}}\x\vec{d}_{2,0}^{\mathrm{M}}\right)\t\vec{d}_{1,2}^{\mathrm{M}}\right]$, and the field pseudovector $\left[\tilde{\vec{\mathcal{E}}}^{\mathrm{L}*}\left(\omega_{1,0}\right)\x\tilde{\vec{\mathcal{E}}}^{\mathrm{L}}\left(\omega_{2,0}\right)\right]$.

Note that before the averaging we had scalar products of dipoles and fields [see Eq. \eqref{eq:metadipole1}].
The averaging over the distribution of  $M$-states in  Eq. \eqref{eq:metadipole1} plays the same role as  
averaging over a random classical rotational ensemble in Eq. \eqref{eq:PXCD_general}: it leads to rearrangement of terms and to the appearance of a rotationally invariant molecular pseudoscalar. 
It shows the link to
the PXCD effect \cite{beaulieu_PXCD_2016, *beaulieu_PXCD} in the vibronic states. 

Eq. \eqref{eq:d_chi_rotational_states} is applicable for an arbitrary field configuration. 
In the work by Patterson et al. \cite{patterson_sensitive_2013}
two linearly polarized fields, orthogonal to each other, have been
used to produce a sum-frequency signal polarized along the direction
perpendicular to both fields. Here we derived the complementary difference-frequency signal.

Importantly, our  result shows that, if two different 
pulses are used, the signal in Eqs. \eqref{eq:PXCD_general}
and \eqref{eq:d_chi_rotational_states} depends on the relative phase between the two pulses. Therefore, the chiral signal 
will only be observed in a reproducible fashion if the 
relative phase between the two pulses is stable
from shot to shot. Clearly, this
is automatically satisfied in case of one-pulse excitation with a
circularly polarized field, where the relative phase between the two perpendicular components is fixed at $\pi/2$, as it happens in PXCD. 

Now we shall consider an alternative scheme, invented by Patterson et al. and involving a static field \cite{patterson_enantiomer-specific_2013}. 

\subsubsection{Vectorial formulation for the static field case}\label{sub:AppendixStatic}

Consider a molecule with eigenstates $\ket n$ in the absence of fields
and initially in the state $\ket 0.$ Application of a static field
$\vec{E}_{S}^{\mathrm{L}}$ transforms the zeroth-order eigenstates
into

\begin{equation}
\ket{n^{\prime}}=\ket n+\sum_{m\neq n}\frac{\vec{E}_{S}^{\mathrm{L}}\t\vec{d}_{m,n}^{\mathrm{L}}}{E_{m,n}}\ket m,
\end{equation}

where $E_{m,n}$ is the energy difference between the $m$-th and
$n$-th states, and we assumed that the states are non-degenerate, or that the perturbation does not couple degenerate states with the same energy. If the perturbation of the initial state is much smaller
than that of the excited state and we apply an oscillating field resonant
with the transition $\ket 0\rightarrow\ket{n^{\prime}}$, then the
first order (in the oscillating field) amplitude of the state $\ket{n^{\prime}}$
reads as

\begin{eqnarray}
a_{n^{\prime}} & = & \i\left[\vec{d}_{n^{\prime},0}^{\mathrm{L}}\t\tilde{\vec{\mathcal{E}}}^{\mathrm{L}}\left(\omega_{n^{\prime},0}\right)\right]\nonumber \\
 & = & \i\left\{ \left[\vec{d}_{n,0}^{\mathrm{L}}\t\tilde{\vec{\mathcal{E}}}^{\mathrm{L}}\left(\omega_{n^{\prime},0}\right)\right]+\sum_{m\neq n}\frac{\vec{E}_{S}^{\mathrm{L}}\t\vec{d}_{n,m}^{\mathrm{L}}}{E_{m,n}}\left[\vec{d}_{m,0}^{\mathrm{L}}\t\tilde{\vec{\mathcal{E}}}^{\mathrm{L}}\left(\omega_{n^{\prime},0}\right)\right]\right\} .
\end{eqnarray}

While the DC Stark field is still present, the expected value of the dipole
has the form

\begin{equation}
\left\langle \vec{d}^{\mathrm{L}}\right\rangle =\vec{d}_{0,0}^{\mathrm{L}}+\left|a_{n^{\prime}}\right|^{2}\vec{d}_{n^{\prime},n^{\prime}}^{\mathrm{L}}+\left(a_{n^{\prime}}\vec{d}_{0,n^{\prime}}^{\mathrm{L}}\e^{-\i\omega_{n^{\prime},0}t}+\mathrm{c.c.}\right).
\end{equation}

Upon orientation averaging, the oscillating term reads as 

\begin{eqnarray}
\int\mathrm{d}\varrho a_{n^{\prime}}\vec{d}_{0,n^{\prime}}^{\mathrm{L}} \e^{-\i\omega_{n^{\prime},0}t}+\mathrm{c.c.} & = & \i\int\mathrm{d}\varrho\left[\vec{d}_{n^{\prime},0}^{\mathrm{L}}\t\tilde{\vec{\mathcal{E}}}^{\mathrm{L}}\left(\omega_{n^{\prime},0}\right)\right]\vec{d}_{0,n^{\prime}}^{\mathrm{L}} \e^{-\i\omega_{n^{\prime},0}t}+\mathrm{c.c.} \nonumber \\
 & = & \frac{\i}{3}\left[\vec{d}_{n^{\prime},0}^{\mathrm{M}}\t\vec{d}_{0,n^{\prime}}^{\mathrm{M}}\right]\tilde{\vec{\mathcal{E}}}^{\mathrm{L}}\left(\omega_{n^{\prime},0}\right) \e^{-\i\omega_{n^{\prime},0}t}+\mathrm{c.c.} ,\label{eq:static_field_achiral}
\end{eqnarray}

so that the oscillations of the induced polarization follow the field.
Note that the orientation averaging for the rotational states would
follow accordingly as shown above,
by replacing $\int\mathrm{d}\varrho$ by a sum over all $M$'s and
keeping the sum on the right hand side of Eq. \eqref{eq:static_field_achiral}.

On the other hand, if the static field is adiabatically removed so
that all of the population in state $\ket{n^{\prime}}$ is transferred
to state $\ket n$ we get

\begin{equation}
\left\langle \vec{d}^{\mathrm{L}}\right\rangle =\vec{d}_{0,0}^{\mathrm{L}}+\left|a_{n^{\prime}}\right|^{2}\vec{d}_{n,n}^{\mathrm{L}}+\left(a_{n^{\prime}}\vec{d}_{0,n}^{\mathrm{L}}\e^{-\i\omega_{n,0}t+\phi}+\mathrm{c.c.}\right)
\end{equation}

where $\phi$ depends on the details of the turn-off of the static field.  The orientation-averaged oscillating term reads as

\begin{eqnarray}
&&\int\mathrm{d}\varrho\,a_{n^{\prime}}\vec{d}_{0,n}^{\mathrm{L}}\e^{-\i\omega_{n,0}t+\phi}+\mathrm{c.c.}\nonumber\\
&=&\i\int\mathrm{d}\varrho\bigg\{\left[\vec{d}_{n,0}^{\mathrm{L}}\t\tilde{\vec{\mathcal{E}}}^{\mathrm{L}}\left(\omega_{n^{\prime},0}\right)\right]\vec{d}_{0,n}^{\mathrm{L}}\nonumber\\
&&+\sum_{m\neq n}\frac{1}{E_{m,n}}\left[\vec{d}_{n,m}^{\mathrm{L}}\t\vec{E}_{S}^{\mathrm{L}}\right]\left[\vec{d}_{m,0}^{\mathrm{L}}\t\tilde{\vec{\mathcal{E}}}^{\mathrm{L}}\left(\omega_{n^{\prime},0}\right)\right]\vec{d}_{0,n}^{\mathrm{L}}\bigg\}\e^{-\i\omega_{n,0}t+\phi}+\mathrm{c.c.}\nonumber\\
&=&\i\bigg\{\frac{1}{3}\left[\vec{d}_{n,0}^{\mathrm{M}}\t\vec{d}_{0,n}^{\mathrm{M}}\right]\tilde{\vec{\mathcal{E}}}^{\mathrm{L}}\left(\omega_{n^{\prime},0}\right)\nonumber\\
&&+\frac{1}{6}\sum_{m\neq n}\frac{1}{E_{m,n}}\left[\left(\vec{d}_{n,m}^{\mathrm{M}}\x\vec{d}_{m,0}^{\mathrm{M}}\right)\t\vec{d}_{0,n}^{\mathrm{M}}\right]\left[\vec{E}_{S}^{\mathrm{L}}\x\tilde{\vec{\mathcal{E}}}^{\mathrm{L}}\left(\omega_{n^{\prime},0}\right)\right]\bigg\}\e^{-\i\omega_{n,0}t+\phi}+\mathrm{c.c.},\label{eq:static}
\end{eqnarray}

In this case we obtain an enantio-sensitive contribution
which oscillates in the direction specified by the cross product between
the direction of the static field and the polarization of the oscillating
field. If, like in the original experiment \cite{patterson_enantiomer-specific_2013},
the static field is along $\hat{x}$ and the oscillating field is
along $\hat{z}$, then the polarization will exhibit oscillations
along $\hat{y}$.

Wave mixing phenomena are usually described on the language of susceptibilities. The quadratic susceptibility  $\chi^{(2)}$ is responsible for three wave mixing. However, both PXCD and EMWS can also be described as free induction decay. In fact, PXCD maximizes  when the laser field is already turned off (see Fig. 2b in Ref. \cite{beaulieu_PXCD_2016, *beaulieu_PXCD}), supporting that free induction decay after the pulse is at its main origin.  

The example of a static field is interesting because it shows that the free induction decay occurring both in PXCD and in EMWS can have very different properties from the ``instantaneous'' response of an isotropic chiral medium described by the quadratic susceptibility $\chi^{(2)}$.
For example, as shown in \cite{fischer2001isotropic}, the chiral quadratic susceptibility vanishes if one of the excitation fields is static, while the second term in Eq. \eqref{eq:static} shows that the chiral response associated with the free induction decay is non-zero, be it EMWS or generalized PXCD. 

\subsection{Bound-bound + bound-unbound transition\label{sub:Generalized PXECD}}

In the previous section we saw how molecular chirality can be
read out from the dynamics of the induced polarization. One can also
imagine reading out this chirality not by looking at the induced polarization
directly but by looking at the photoelectron current induced by a
second absorption process as originally proposed in \cite{beaulieu_PXCD_2016, *beaulieu_PXCD}.
Here, we will consider the general case in which a pump pulse of arbitrary
polarization excites the molecule to a bound superposition and a probe
pulse of arbitrary polarization ionizes it after a time delay $\tau$.
In this case the photoionization amplitude into the state $\vert \vec{k}^{\mathrm{M}}\rangle$ reads as 

\begin{eqnarray}
a_{\vec{k}^{\mathrm{M}}} & = & -\left[\vec{d}_{1,0}^{\mathrm{L}}\t\tilde{\vec{\mathcal{E}}}_{1}^{\mathrm{L}}\left(\omega_{1,0}\right)\right]\left[\vec{D}_{1}^{\mathrm{L}}\t\tilde{\vec{\mathcal{E}}}_{2}^{\mathrm{L}}\left(\omega_{k,1}\right)\right]\e^{-\i\omega_{1}\tau}\nonumber \\
 &  & -\left[\vec{d}_{2,0}^{\mathrm{L}}\t\tilde{\vec{\mathcal{E}}}_{1}^{\mathrm{L}}\left(\omega_{2,0}\right)\right]\left[\vec{D}_{2}^{\mathrm{L}}\t\tilde{\vec{\mathcal{E}}}_{2}^{\mathrm{L}}\left(\omega_{k,2}\right)\right]\e^{-\i\omega_{2}\tau},\label{eq:a_k_twophoton}
\end{eqnarray}

where $\vec{d}_{i,0}$ is a bound-bound transition dipole between states $\vert i \rangle$ and $\vert 0 \rangle$, $\vec{D}_i$ is a bound-continuum transition dipole between states $\vert \vec{k}^{\mathrm{M}} \rangle$ and $\vert i \rangle$, $\tilde{\vec{\mathcal{E}}}_i$ is the Fourier transform of the $i$-th pulse, and we assumed that the pulses do not overlap. Application of Eq.
\eqref{eq:5th_rank_tensor_invariant_vector_identity} to Eqs. \eqref{eq:j_mol},
\eqref{eq:j_lab_total}, and \eqref{eq:a_k_twophoton}, yields the
most general result and it shows that not only the cross terms, as
in the generalized PXCD {[}see Eq. \eqref{eq:PXCD_general}{]}, but
also the diagonal terms in $\vert a_{\vec{k}^{\mathrm{M}}}\vert^{2}$
may contribute to the net photoelectron current

\begin{eqnarray}
\vec{j}^{\mathrm{L}}\left(k\right) &=& \int \mathrm{d} \varrho \int \mathrm{d} \Omega_{k}^\mathrm{M} \vert a_{\vec{k}^\mathrm{M}} (\varrho)\vert^{2} S(\varrho)\vec{k}^{\mathrm{M}} \nonumber \\
&=&\vec{j}_{\mathrm{diag},1}^{\mathrm{L}}\left(k\right)+\vec{j}_{\mathrm{diag},2}^{\mathrm{L}}\left(k\right)+\vec{j}_{\mathrm{cross}}^{\mathrm{L}}\left(k\right).\label{eq:j_PXECD}
\end{eqnarray}

The contribution from the diagonal terms is of the form

\begin{eqnarray}
\vec{j}_{\mathrm{diag},i}^{\mathrm{L}}\left(k\right) & = & \int\mathrm{d}\varrho \int \mathrm{d} \Omega_{k}^\mathrm{M}\,\left(\vec{d}_{0,i}^{\mathrm{L}}\t\tilde{\vec{\mathcal{E}}}_{1}^{\mathrm{L}*}\right)\left(\vec{D}_{i}^{\mathrm{L}*}\t\tilde{\vec{\mathcal{E}}}_{2}^{\mathrm{L}*}\right)\left(\vec{d}_{i,0}^{\mathrm{L}}\t\tilde{\vec{\mathcal{E}}}_{1}^{\mathrm{L}}\right)\left(\vec{D}_{i}^{\mathrm{L}}\t\tilde{\vec{\mathcal{E}}}_{2}^{\mathrm{L}}\right)\vec{k}^{\mathrm{L}}\nonumber \\
 & = & \frac{1}{15}\Re\bigg\{\int\mathrm{d}\Omega_{k}^{\mathrm{M}}\left[\left(\vec{d}_{0,i}^{\mathrm{M}}\x\vec{D}_{i}^{\mathrm{M}*}\right)\t\vec{D}_{i}^{\mathrm{M}}\right]\left(\vec{d}_{i,0}^{\mathrm{M}}\t\vec{k}^{\mathrm{M}}\right)\left[\left(\tilde{\vec{\mathcal{E}}}_{1}^{\mathrm{L}*}\x\tilde{\vec{\mathcal{E}}}_{2}^{\mathrm{L}*}\right)\t\tilde{\vec{\mathcal{E}}}_{2}^{\mathrm{L}}\right]\tilde{\vec{\mathcal{E}}}_{1}^{\mathrm{L}}\nonumber \\
 &  & +\int\mathrm{d}\Omega_{k}^{\mathrm{M}}\left[\left(\vec{d}_{0,i}^{\mathrm{M}}\x\vec{D}_{i}^{\mathrm{M}*}\right)\t\vec{k}^{\mathrm{M}}\right]\left(\vec{d}_{i,0}^{\mathrm{M}}\t\vec{D}_{i}^{\mathrm{M}}\right)\left(\tilde{\vec{\mathcal{E}}}_{1}^{\mathrm{L}}\t\tilde{\vec{\mathcal{E}}}_{2}^{\mathrm{L}}\right)\left(\tilde{\vec{\mathcal{E}}}_{1}^{\mathrm{L}*}\x\tilde{\vec{\mathcal{E}}}_{2}^{\mathrm{L}*}\right)\nonumber \\
 &  & +\int\mathrm{d}\Omega_{k}^{\mathrm{M}}\left[\left(\vec{d}_{0,i}^{\mathrm{M}}\x\vec{D}_{i}^{\mathrm{M}}\right)\t\vec{k}^{\mathrm{M}}\right]\left(\vec{d}_{i,0}^{\mathrm{M}}\t\vec{D}_{i}^{\mathrm{M}*}\right)\left(\tilde{\vec{\mathcal{E}}}_{1}^{\mathrm{L}}\t\tilde{\vec{\mathcal{E}}}_{2}^{\mathrm{L}*}\right)\left(\tilde{\vec{\mathcal{E}}}_{1}^{\mathrm{L}*}\x\tilde{\vec{\mathcal{E}}}_{2}^{\mathrm{L}}\right)\bigg\}\nonumber \\
 &  & +\frac{1}{30}\left|\vec{d}_{i,0}^{\mathrm{M}}\right|^{2}\left|\tilde{\vec{\mathcal{E}}}_{1}^{\mathrm{L}}\right|^{2}\int\mathrm{d}\Omega_{k}^{\mathrm{M}}\left[\left(\vec{D}_{i}^{\mathrm{M}*}\x\vec{D}_{i}^{\mathrm{M}}\right)\t\vec{k}^{\mathrm{M}}\right]\left(\tilde{\vec{\mathcal{E}}}_{2}^{\mathrm{L}*}\x\tilde{\vec{\mathcal{E}}}_{2}^{\mathrm{L}}\right)\label{eq:j_diag}
\end{eqnarray}

where we only assumed that $\vec{d}_{0,i}=\vec{d}_{i,0}$ is real, which
can always be achieved for bound states in the absence of magnetic
fields. The fields $\tilde{\vec{\mathcal{E}}}_{1}$ and $\tilde{\vec{\mathcal{E}}}_{2}$
are evaluated at the frequencies $\omega_{i,0}$ and $\omega_{k,i}$
respectively. The last term is simply the generalized PECD from the
$i$-th state multiplied by the population in the $i$-th state induced
by the pump and a factor of $1/5$ that comes from the orientation averaging.
The terms in curly brackets represent contributions
to the current beyond the usual PECD. Each term has selection rules
that are evident from its vectorial structure, and will be discussed
below after considering the cross terms contribution to the photoelectron
current. As usual, the molecular terms are rotationally-invariant molecule-specific pseudoscalars and the
field terms are pseudovectors. 

The contribution from the cross terms in $\vert a_{\vec{k}^{\mathrm{M}}}\vert^{2}$
to the net photoelectron current $\vec{j}^{\mathrm{L}}\left(k\right)$
is given in general by

\begin{eqnarray}
\vec{j}_{\mathrm{cross}}^{\mathrm{L}}\left(k\right) & = & \int\mathrm{d}\varrho \int \mathrm{d} \Omega_{k}^\mathrm{M}\,\left(\vec{d}_{0,2}^{\mathrm{L}}\t\tilde{\vec{\mathcal{E}}}_{1}^{\mathrm{L}*}\right)\left(\vec{D}_{2}^{\mathrm{L}*}\t\tilde{\vec{\mathcal{E}}}_{2}^{\mathrm{L}*}\right)\left(\vec{d}_{1,0}^{\mathrm{L}}\t\tilde{\vec{\mathcal{E}}}_{1}^{\mathrm{L}}\right)\left(\vec{D}_{1}^{\mathrm{L}}\t\tilde{\vec{\mathcal{E}}}_{2}^{\mathrm{L}}\right)\vec{k}^{\mathrm{L}}+\mathrm{c.c.}\nonumber \\
 & = & \vec{j}_{\mathrm{noncopl}}^{\mathrm{L}}\left(k\right)+\vec{j}_{\mathrm{ellip}}^{\mathrm{L}}\left(k\right)+\vec{j}_{\mathrm{lin}}^{\mathrm{L}}\left(k\right),\label{eq:j_cross}
\end{eqnarray}

where the fields $\tilde{\vec{\mathcal{E}}}_{1}^{\mathrm{L}*}$, $\tilde{\vec{\mathcal{E}}}_{2}^{\mathrm{L}*}$,
$\tilde{\vec{\mathcal{E}}}_{1}^{\mathrm{L}}$, and $\tilde{\vec{\mathcal{E}}}_{2}^{\mathrm{L}}$
are evaluated at the frequencies $\omega_{2,0}$, $\omega_{k,2}$,
$\omega_{1,0}$, and $\omega_{k,1}$, respectively, and we grouped
the 10 terms according to their selection rules for the fields as
follows. The first group reads as

\begin{eqnarray}
\vec{j}_{\mathrm{noncopl}}^{\mathrm{L}}\left(k\right) & = & \frac{1}{30}\bigg\{\int\mathrm{d}\Omega_{k}^{\mathrm{M}}\left[\left(\vec{d}_{0,2}^{\mathrm{M}}\x\vec{D}_{2}^{\mathrm{M}*}\right)\t\vec{d}_{1,0}^{\mathrm{M}}\right]\left(\vec{D}_{1}^{\mathrm{M}}\t\vec{k}^{\mathrm{M}}\right)\left[\left(\tilde{\vec{\mathcal{E}}}_{1}^{\mathrm{L}*}\x\tilde{\vec{\mathcal{E}}}_{2}^{\mathrm{L}*}\right)\t\tilde{\vec{\mathcal{E}}}_{1}^{\mathrm{L}}\right]\tilde{\vec{\mathcal{E}}}_{2}^{\mathrm{L}}\nonumber \\
 &  & +\int\mathrm{d}\Omega_{k}^{\mathrm{M}}\left[\left(\vec{d}_{0,2}^{\mathrm{M}}\x\vec{D}_{2}^{\mathrm{M}*}\right)\t\vec{D}_{1}^{\mathrm{M}}\right]\left(\vec{d}_{1,0}^{\mathrm{M}}\t\vec{k}^{\mathrm{M}}\right)\left[\left(\tilde{\vec{\mathcal{E}}}_{1}^{\mathrm{L}*}\x\tilde{\vec{\mathcal{E}}}_{2}^{\mathrm{L}*}\right)\t\tilde{\vec{\mathcal{E}}}_{2}^{\mathrm{L}}\right]\tilde{\vec{\mathcal{E}}}_{1}^{\mathrm{L}}\nonumber \\
 &  & +\int\mathrm{d}\Omega_{k}^{\mathrm{M}}\left[\left(\vec{d}_{0,2}^{\mathrm{M}}\x\vec{d}_{1,0}^{\mathrm{M}}\right)\t\vec{D}_{1}^{\mathrm{M}}\right]\left(\vec{D}_{2}^{\mathrm{M}*}\t\vec{k}^{\mathrm{M}}\right)\left[\left(\tilde{\vec{\mathcal{E}}}_{1}^{\mathrm{L}*}\x\tilde{\vec{\mathcal{E}}}_{1}^{\mathrm{L}}\right)\t\tilde{\vec{\mathcal{E}}}_{2}^{\mathrm{L}}\right]\tilde{\vec{\mathcal{E}}}_{2}^{\mathrm{L}*}\nonumber \\
 &  & +\int\mathrm{d}\Omega_{k}^{\mathrm{M}}\left[\left(\vec{D}_{2}^{\mathrm{M}*}\x\vec{d}_{1,0}^{\mathrm{M}}\right)\t\vec{D}_{1}^{\mathrm{M}}\right]\left(\vec{d}_{0,2}^{\mathrm{M}}\t\vec{k}^{\mathrm{M}}\right)\left[\left(\tilde{\vec{\mathcal{E}}}_{2}^{\mathrm{L}*}\x\tilde{\vec{\mathcal{E}}}_{1}^{\mathrm{L}}\right)\t\tilde{\vec{\mathcal{E}}}_{2}^{\mathrm{L}}\right]\tilde{\vec{\mathcal{E}}}_{1}^{\mathrm{L}*}\bigg\}\e^{\i\omega_{21}\tau}\nonumber \\
 &  & +\mathrm{c.c.}, \label{eq:j_noncopl}
\end{eqnarray}

and contains all the terms involving scalar triple products of the field vectors, which means that each of
its terms vanishes if the fields involved in its triple product are
coplanar. It means that exciting $\vec{j}_{\mathrm{noncopl}}^{\mathrm{L}}\left(k\right)$ requires non-collinear geometry of pump and probe pulses. For fields with the same polarization at the two transition
frequencies, that is, $\tilde{\vec{\mathcal{E}}}_{1}^{\mathrm{L}}\left(\omega_{1,0}\right)\parallel\tilde{\vec{\mathcal{E}}}_{1}^{\mathrm{L}}\left(\omega_{2,0}\right)$
and $\tilde{\vec{\mathcal{E}}}_{2}^{\mathrm{L}}\left(\omega_{k,1}\right)\parallel\tilde{\vec{\mathcal{E}}}_{2}^{\mathrm{L}}\left(\omega_{k,2}\right)$,
$\vec{j}_{\mathrm{noncopl}}^{\mathrm{L}}$ vanishes unless the polarization
of the pump and the probe are non-coplanar, which means that at least
one of the fields must be elliptically polarized. The other field
can be either linearly or elliptically polarized, provided its polarization
is non-coplanar to that of the first field. 

The second group of contributions to $\vec{j}_{\mathrm{cross}}^{\mathrm{L}}$
is given by

\begin{eqnarray}
\vec{j}_{\mathrm{ellip}}^{\mathrm{L}}\left(k\right) & = & \frac{1}{30}\bigg\{\int\mathrm{d}\Omega_{k}^{\mathrm{M}}\left[\left(\vec{d}_{0,2}^{\mathrm{M}}\x\vec{d}_{1,0}^{\mathrm{M}}\right)\t\vec{k}^{\mathrm{M}}\right]\left(\vec{D}_{2}^{\mathrm{M}*}\t\vec{D}_{1}^{\mathrm{M}}\right)\left(\tilde{\vec{\mathcal{E}}}_{2}^{\mathrm{L}*}\t\tilde{\vec{\mathcal{E}}}_{2}^{\mathrm{L}}\right)\left(\tilde{\vec{\mathcal{E}}}_{1}^{\mathrm{L}*}\x\tilde{\vec{\mathcal{E}}}_{1}^{\mathrm{L}}\right)\nonumber \\
 &  & +\int\mathrm{d}\Omega_{k}^{\mathrm{M}}\left[\left(\vec{D}_{2}^{\mathrm{M}*}\x\vec{D}_{1}^{\mathrm{M}}\right)\t\vec{k}^{\mathrm{M}}\right]\left(\vec{d}_{0,2}^{\mathrm{M}}\t\vec{d}_{1,0}^{\mathrm{M}}\right)\left(\tilde{\vec{\mathcal{E}}}_{1}^{\mathrm{L}*}\t\tilde{\vec{\mathcal{E}}}_{1}^{\mathrm{L}}\right)\left(\tilde{\vec{\mathcal{E}}}_{2}^{\mathrm{L}*}\x\tilde{\vec{\mathcal{E}}}_{2}^{\mathrm{L}}\right)\bigg\}\e^{\i\omega_{21}\tau}\nonumber \\
 &  & +\mathrm{c.c.}\label{eq:j_ellip}
\end{eqnarray}

and contains the two terms involving a cross product between a single
field at the two transition frequencies. For fields satisfying $\tilde{\vec{\mathcal{E}}}_{1}^{\mathrm{L}}\left(\omega_{1,0}\right)\parallel\tilde{\vec{\mathcal{E}}}_{1}^{\mathrm{L}}\left(\omega_{2,0}\right)$
and $\tilde{\vec{\mathcal{E}}}_{2}^{\mathrm{L}}\left(\omega_{k,1}\right)\parallel\tilde{\vec{\mathcal{E}}}_{2}^{\mathrm{L}}\left(\omega_{k,2}\right)$,
each term vanishes unless the field in the cross product is elliptically
polarized. The field in the scalar product can have any polarization. 

The third group of contributions to $\vec{j}_{\mathrm{cross}}^{\mathrm{L}}$
reads as

\begin{eqnarray}
\vec{j}_{\mathrm{lin}}^{\mathrm{L}}\left(k\right) & = & \frac{1}{30}\bigg\{\int\mathrm{d}\Omega_{k}^{\mathrm{M}}\left[\left(\vec{d}_{0,2}^{\mathrm{M}}\x\vec{D}_{2}^{\mathrm{M}*}\right)\t\vec{k}^{\mathrm{M}}\right]\left(\vec{d}_{1,0}^{\mathrm{M}}\t\vec{D}_{1}^{\mathrm{M}}\right)\left(\tilde{\vec{\mathcal{E}}}_{1}^{\mathrm{L}}\t\tilde{\vec{\mathcal{E}}}_{2}^{\mathrm{L}}\right)\left(\tilde{\vec{\mathcal{E}}}_{1}^{\mathrm{L}*}\x\tilde{\vec{\mathcal{E}}}_{2}^{\mathrm{L}*}\right)\nonumber \\
 &  & +\int\mathrm{d}\Omega_{k}^{\mathrm{M}}\left[\left(\vec{d}_{0,2}^{\mathrm{M}}\x\vec{D}_{1}^{\mathrm{M}}\right)\t\vec{k}^{\mathrm{M}}\right]\left(\vec{D}_{2}^{\mathrm{M}*}\t\vec{d}_{1,0}^{\mathrm{M}}\right)\left(\tilde{\vec{\mathcal{E}}}_{2}^{\mathrm{L}*}\t\tilde{\vec{\mathcal{E}}}_{1}^{\mathrm{L}}\right)\left(\tilde{\vec{\mathcal{E}}}_{1}^{\mathrm{L}*}\x\tilde{\vec{\mathcal{E}}}_{2}^{\mathrm{L}}\right)\nonumber \\
 &  & +\int\mathrm{d}\Omega_{k}^{\mathrm{M}}\left[\left(\vec{D}_{2}^{\mathrm{M}*}\x\vec{d}_{1,0}^{\mathrm{M}}\right)\t\vec{k}^{\mathrm{M}}\right]\left(\vec{d}_{0,2}^{\mathrm{M}}\t\vec{D}_{1}^{\mathrm{M}}\right)\left(\tilde{\vec{\mathcal{E}}}_{1}^{\mathrm{L}*}\t\tilde{\vec{\mathcal{E}}}_{2}^{\mathrm{L}}\right)\left(\tilde{\vec{\mathcal{E}}}_{2}^{\mathrm{L}*}\x\tilde{\vec{\mathcal{E}}}_{1}^{\mathrm{L}}\right)\nonumber \\
 &  & +\int\mathrm{d}\Omega_{k}^{\mathrm{M}}\left[\left(\vec{d}_{1,0}^{\mathrm{M}}\x\vec{D}_{1}^{\mathrm{M}}\right)\t\vec{k}^{\mathrm{M}}\right]\left(\vec{d}_{0,2}^{\mathrm{M}}\t\vec{D}_{2}^{\mathrm{M}*}\right)\left(\tilde{\vec{\mathcal{E}}}_{1}^{\mathrm{L}*}\t\tilde{\vec{\mathcal{E}}}_{2}^{\mathrm{L}*}\right)\left(\tilde{\vec{\mathcal{E}}}_{1}^{\mathrm{L}}\x\tilde{\vec{\mathcal{E}}}_{2}^{\mathrm{L}}\right)\bigg\}\e^{\i\omega_{21}\tau}\nonumber \\
 &  & +\mathrm{c.c.}\label{eq:j_lin}
\end{eqnarray}

and contains the remaining terms. Unlike $\vec{j}_{\mathrm{noncopl}}^{\mathrm{L}}$
and $\vec{j}_{\mathrm{ellip}}^{\mathrm{L}}$, which vanish in the
absence of elliptical fields when $\tilde{\vec{\mathcal{E}}}_{1}^{\mathrm{L}}\left(\omega_{1,0}\right)\parallel\tilde{\vec{\mathcal{E}}}_{1}^{\mathrm{L}}\left(\omega_{2,0}\right)$
and $\tilde{\vec{\mathcal{E}}}_{2}^{\mathrm{L}}\left(\omega_{k,1}\right)\parallel\tilde{\vec{\mathcal{E}}}_{2}^{\mathrm{L}}\left(\omega_{k,2}\right)$,
$\vec{j}_{\mathrm{lin}}^{\mathrm{L}}$ can be non-zero even for purely
linear fields provided pump and probe are neither parallel nor orthogonal to
each other. Clearly, the selection rules described for $\vec{j}_{\mathrm{noncopl}}^{\mathrm{L}}$,
$\vec{j}_{\mathrm{lin}}^{\mathrm{L}}$, and $\vec{j}_{\mathrm{ellip}}^{\mathrm{L}}$
are also valid for the first, the second and third, and the last term
of $\vec{j}_{\mathrm{diag}}^{\mathrm{L}}$, respectively.

As a whole, the 10 terms in Eq. \eqref{eq:j_cross} correspond to
the 10 ways in which the five molecular vectors $\vec{d}_{0,1}^{\mathrm{M}}$,
$\vec{d}_{2,0}^{\mathrm{M}}$, $\vec{D}_{1}^{\mathrm{M}}$, $\vec{D}_{2}^{\mathrm{M}}$,
and $\vec{k}^{\mathrm{M}}$ can form a rotation-invariant molecular
quantity. Each molecular term is coupled to a field term that corresponds
to 1 of the 10 ways that a vector can be formed via scalar and vector
products between 4 vectors. Unlike the diagonal terms, the cross terms
contribution oscillates with the pump-probe time delay at a frequency
corresponding to the energy difference between the two bound states
excited by the pump.

If we consider the PXECD setup originally described in \cite{beaulieu_PXCD_2016, *beaulieu_PXCD},
where the pump field is circularly polarized like in Eq. \eqref{eq:field}
and the pump is linearly polarized along $\hat{x}^{\mathrm{L}}$, then application of the above discussed selection rules and some vector algebra (see Appendix \ref{sub:AppendixCircLin}) yields

\begin{equation}
\vec{j}^{\mathrm{L}}\left(k\right)=\frac{\i\sigma}{60}\tilde{\mathcal{E}}_{1}^{*}\tilde{\mathcal{E}}_{2}^{*}\tilde{\mathcal{E}}_{1}\tilde{\mathcal{E}}_{2}\left[\left(\vec{d}_{0,1}^{\mathrm{M}}\x\vec{d}_{2,0}^{\mathrm{M}}\right)\t\int\mathrm{d}\Omega_{k}^{\mathrm{M}}\vec{D}_{12}^{\mathrm{M}}\left(\vec{k}^{\mathrm{M}}\right)\right]\hat{z}^{\mathrm{L}}\e^{\i\omega_{21}\tau}+\mathrm{c.c.},\label{eq:j_circ+linear}
\end{equation}

\begin{equation}
\vec{D}_{12}^{\mathrm{M}}\left(\vec{k}^{\mathrm{M}}\right)=-4\left(\vec{D}_{1}^{\mathrm{M}}\t\vec{D}_{2}^{\mathrm{M}*}\right)\vec{k}^{\mathrm{M}}+\left(\vec{D}_{2}^{\mathrm{M}*}\t\vec{k}^{\mathrm{M}}\right)\vec{D}_{1}^{\mathrm{M}}+\left(\vec{D}_{1}^{\mathrm{M}}\t\vec{k}^{\mathrm{M}}\right)\vec{D}_{2}^{\mathrm{M}*},\label{eq:D12_circ_linear}
\end{equation}

which coincides with the result originally obtained in \cite{beaulieu_PXCD_2016, *beaulieu_PXCD}. Eqs. \eqref{eq:j_PXECD}, \eqref{eq:j_diag}, \eqref{eq:j_cross}, \eqref{eq:j_noncopl}, \eqref{eq:j_ellip}, and \eqref{eq:j_lin} are the generalization of PXECD to arbitrary polarizations of the pump and probe pulses. 

Interestingly, although the symmetry of a linear pump - linear probe
scheme where the two fields are orthogonal to each other does not
forbid the emergence of a non-zero net photoelectron current $\vec{j}^{\mathrm{L}}$
(see Fig. \ref{fig:Symmetry_xy}), Eqs. \eqref{eq:j_PXECD}, \eqref{eq:j_diag}, \eqref{eq:j_cross}, \eqref{eq:j_noncopl}, \eqref{eq:j_ellip}, and \eqref{eq:j_lin}
show that it vanishes. This symmetry can be traced back to the fact that the phase shift between the pump and the probe is not recorded by the system 
because the probe step corresponds to the parametric process in terms of non-linear optics diagrams (see Fig. 1 in Ref. \cite{beaulieu_PXCD_2016, *beaulieu_PXCD}), where the initial and final states are the same: it is a superposition of the states prepared by the pump.
It highlights the fact that all the effects considered in this section do not require a phase-lock between the pump and probe pulses. 

\section{Conclusions}\label{sec:Conclusions}

We have presented a unified approach to electric-dipole-based methods of chiral discrimination. The approach is based on a vectorial formulation of the chiral response and provides a common language for understanding electric-dipole-based techniques used in different fields, such as photoionization and microwave spectroscopy. 
All these techniques make use of coherent excitation of several states leading to electronic, vibronic, rotational, or ionization dynamics.

The chiral response in all cases is characterized by a vectorial observable and takes place within a chiral setup. Unlike scalar observables (e.g. total cross sections), vectorial observables (e.g. induced polarization) are able to exploit the chirality of such setups and therefore provide the opportunity to probe the chirality of isotropic molecular samples without relying on the chirality of the light inducing the response.
Chiral setups can result from the combination of at least two linearly polarized fields with non-collinear polarizations (and phase-delayed in the case of a single frequency) defining a non-zero pseudovector, and a detector defining a direction parallel or anti-parallel to the field pseudovector. Furthermore, the fields defining the pseudovector need not overlap in time, which allows for pump-probe schemes in the construction of the chiral setup.

We have shown that the generic structure of the vectorial observable is given by the product of the field pseudovector, defined by the configurations of the electric fields exciting or probing chiral dynamics, and a molecular pseudoscalar characterizing the molecular handedness. The projection of the vectorial observable on the 
\emph{external} direction defined by the detector yields the result of the measurement: a product between the molecular pseudoscalar associated to the molecular handedness, and the chiral setup pseudoscalar defining the handedness of the chiral setup. 

The molecular pseudoscalar is given by a rotationally invariant molecule-specific quantity such as 
a triple product involving three bound-bound transition dipoles, and/or the triple product between photoionization dipoles and the photoelectron momentum integrated over all directions. 
The strength of the chiral response is determined by the mutual orientation of such vectors in the molecular frame.  

The affinity of different electric-dipole-based techniques should help us to identify general  mechanisms of chiral response, driven exclusively by the electric component of the electromagnetic field, and their link to molecular chiral structure and dynamics.

\section{Acknowledgements}
O.S. gratefully acknowledges illuminating discussions with Prof. Aephraim Steinberg, in particular,  on the role of the chiral observer in detecting the chiral response in electric-dipole-based methods. We also thank Dr. Emilio Pisanty  for his comments on the role of the chiral setup in PECD. The authors are grateful to Prof. Misha Ivanov for stimulating discussions and comments on the manuscript. 
O.S. thanks Dr. Alex G. Harvey and Dr. Zden\v{e}k Ma\v{s}\'{i}n for useful discussions. 
We thank Dr. Laurent Nahon and Prof. Christiane Koch for their comments on the manuscript and Prof. Amar Vutha for discussions of the EMWS scenarios.
A.F.O. gratefully acknowledges the MEDEA project, which has received funding from the European Union's Horizon 2020 research and innovation programme under the Marie Sk\l{}odowska-Curie grant agreement No 641789. O.S. gratefully acknowledges the QUTIF programme of the Deutsche Forschungsgemeinschaft, project Sm 292-5/1.
\section{Appendix}

\subsection{Beyond the electric-dipole approximation: the magnetic dipole, the
helicity of light, and absorption circular dichroism}\label{sec:AppendixCD}

In order to introduce the reader into some fundamental aspects of the discussion in the main part of the manuscript, we  will briefly
illustrate the relation between magnetic dipole, helicity of light,
and absorption circular dichroism in randomly oriented chiral molecules.

The interaction between the electron in the molecule and the radiation
field can be described by the interaction Hamiltonian (see e.g. \cite{barron_molecular_2004})

\begin{equation}
H^{\prime}\left(t\right)=-\vec{d}\t\vec{E}\left(0,t\right)-\vec{m}\t\vec{B}\left(0,t\right)+\dots\label{eq:-dE-mB}
\end{equation}

where $\vec{d}$ and $\vec{m}$ are the electric and magnetic dipoles, 

\begin{equation}
\vec{E}\left(\vec{r},t\right)=-\partial_{t}\vec{A}\left(\vec{r},t\right)\qquad\mathrm{and}\qquad\vec{B}\left(\vec{r},t\right)=\vec{\nabla}\x\vec{A}\left(\vec{r},t\right)\label{eq:E_and_B}
\end{equation}

are the electric and magnetic fields, and $\vec{A}\left(\vec{r},t\right)$
is the vector potential. Other terms of the same order as the magnetic-dipole
interaction (e.g. the electric-quadrupole interaction) have been
ignored because electric-quadrupole effects vanish in isotropic samples \cite{barron_molecular_2004}. Consider a plane wave
with wave number $\vec{k}$ and frequency $\omega$, 

\begin{equation}
\vec{A}\left(\vec{r},t\right)=\vec{\mathcal{A}}\e^{\i\left(\vec{k}\t\vec{r}-\omega t\right)}+\mathrm{c.c.},\label{eq:vector_potential}
\end{equation}

where $\vec{\mathcal{A}}$ encodes the polarization, intensity, and
phase shift of the wave. For wavelengths $\lambda$ much greater than the electron
orbit, the term $\vec{k}\t\vec{r}=2\pi r/\lambda$ is very small and
$\e^{\i\vec{k}\t\vec{r}}$ can be expanded in powers of it. The electric-dipole
and magnetic-dipole interactions in Eq. \eqref{eq:-dE-mB} stem from
the zeroth and first order terms, respectively, of such expansion.
That is, the magnetic-dipole interaction emerges as a consequence
of taking into account the spatial structure of the electromagnetic
field. Furthermore, absorption circular dichroism, which is linear
in the magnetic-dipole interaction, scales as $\vec{k}\t\vec{r}$,
i.e. as the ratio of the electron orbit size to the wavelength. 

Replacing Eq. \eqref{eq:vector_potential} in Eq. \eqref{eq:E_and_B}
yields 

\begin{equation}
\vec{E}\left(0,t\right)=\vec{\mathcal{E}}\e^{-\i\omega t}+\mathrm{c.c.}\qquad\mathrm{and}\qquad\vec{B}\left(0,t\right)=\vec{\mathcal{B}}\e^{-\i\omega t}+\mathrm{c.c.},
\end{equation}

where $\vec{\mathcal{E}}=\omega\vec{\mathcal{A}}$ and $\vec{\mathcal{B}}=\i\vec{k}\x\vec{\mathcal{A}}$. Therefore, the probability that the molecule in the initial state $\ket i$ is
excited into the upper energy state $\ket f$ is given by

\begin{equation}
\left|\bra fH^{\prime}\left(t\right)\ket i\right|^{2}\propto\left|\left(\vec{d}_{fi}\t\vec{\mathcal{E}}+\vec{m}_{fi}\t\vec{\mathcal{B}}\right)\right|^{2}
\end{equation}

and contains an interference term of the form

\begin{equation}
\left(\vec{d}_{fi}\t\vec{\mathcal{E}}\right)^{*}\left(\vec{m}_{fi}\t\vec{\mathcal{B}}\right)+\mathrm{c.c.}
\end{equation}

For the case of electronic and/or vibrational transitions, $\vec{d}_{fi}$
and $\vec{m}_{fi}$ are fixed in the molecular frame, while $\vec{\mathcal{E}}$
and $\vec{\mathcal{B}}$ are fixed in the lab frame. If the sample
is isotropic we must average over all molecular orientations $\varrho$
(see Appendix \ref{sec:AppendixOrientationAveraging}), which yields

\begin{equation}
\int\mathrm{d}\varrho\,\left[\vec{d}_{fi}^{\mathrm{L}}\left(\varrho\right)\t\vec{\mathcal{E}}^{\mathrm{L}}\right]^{*}\left[\vec{m}_{fi}^{\mathrm{L}}\left(\varrho\right)\t\vec{\mathcal{B}}^{\mathrm{L}}\right]=\frac{1}{3}\left[\vec{d}_{fi}^{\mathrm{M}}\t\vec{m}_{fi}^{\mathrm{M}}\right]\left[\vec{\mathcal{E}}^{\mathrm{L}*}\t\vec{\mathcal{B}}^{\mathrm{L}}\right],\label{eq:CD}
\end{equation}

where the superscripts $\mathrm{L}$ and $\mathrm{M}$ indicate vectors
expressed in the lab and molecular frames respectively, and we explicitly
indicated the dependence of the molecular frame vectors $\vec{d}_{fi}$
and $\vec{m}_{fi}$ on the molecular orientation $\varrho$ when they are expressed
in the lab frame. The right hand side of Eq. \eqref{eq:CD} is a scalar
that is the product of two pseudoscalars. One of them contains only
molecular quantities in the molecular frame, and the other contains
only field quantities in the lab frame. Furthermore, the latter is
proportional to the helicity of the field, i.e. it is proportional
to the projection of the light spin angular momentum on the propagation
direction $\vec{k}$. To see this, we rewrite the field pseudoscalar
in terms of the vector potential as 

\begin{equation}
\vec{\mathcal{E}}^{\mathrm{L}*}\t\vec{\mathcal{B}}^{\mathrm{L}}=\omega\vec{\mathcal{A}}^{\mathrm{L}*}\t\left(\i\vec{k}^{\mathrm{L}}\x\vec{\mathcal{A}}^{\mathrm{L}}\right)=\omega\left(\i\vec{\mathcal{A}}^{\mathrm{L}}\x\mathcal{\vec{A}}^{\mathrm{L}*}\right)\t\vec{k}^{\mathrm{L}}.
\end{equation}

The factor $\i\vec{\mathcal{A}}^{\mathrm{L}}\x\mathcal{\vec{A}}^{\mathrm{L}*}$
is always real and it is proportional to the photon's spin. For example,
for light circularly polarized in the $xy$ plane $\vec{\mathcal{A}}^{\mathrm{L}}=\mathcal{A}\left(\hat{x}^{\mathrm{L}}+\i\sigma\hat{y}^{\mathrm{L}}\right)/\sqrt{2}$,
$\sigma=\pm1$, and $\i\vec{\mathcal{A}}^{\mathrm{L}}\x\mathcal{\vec{A}}^{\mathrm{L}*}=\left|\mathcal{A}\right|^{2}\sigma\hat{z}^{\mathrm{L}}$,
where $\sigma\hat{z}^{\mathrm{L}}$ is the spin of the photon. If
we now project on the propagation direction $\hat{k}^{\mathrm{L}}$, we obtain the sign of the helicity of the circularly polarized field

\begin{equation}
\eta=\sigma\hat{z}^{\mathrm{L}}\t\hat{k}^{\mathrm{L}}=\pm\sigma,
\end{equation}

where we used the fact that $\vec{k}^{\mathrm{L}}$ can point either in the positive (+) or negative (-) $\hat{z}^{\mathrm{L}}$
direction. One must be careful of not confusing $\sigma$ with $\eta$. While
$\eta$ indicates the handedness of the helix formed by the electric
(or magnetic) component of the circularly polarized field in space
at a fixed time and is a time-even pseudoscalar, $\sigma$ merely
indicates the direction of rotation of the electric field in time
at a fixed point in space, is invariant with respect to parity inversion,
and is therefore a time-odd scalar. 

Importantly for the discussion in the main part of the manuscript, in the electric-dipole approximation
the variation of the electromagnetic field in space and along with
it the propagation direction of the light, the magnetic field, and
the magnetic-dipole interaction, are absent. Therefore, the chiral
effects which rely only on the electric-dipole interaction do not
rely on the helicity of the light, but on its spin. In other words,
they do not rely on the pseudoscalar character of the light encoded
in $\eta$ but instead on its time-odd character encoded in the pseudovector
$\sigma\hat{z}$.

\subsection{Classical orientation averaging\label{sec:AppendixOrientationAveraging}}

Following the formalism in Sec. 4.2 of Ref. \cite{barron_molecular_2004}
we can perform the orientation averaging using tensor notation as
follows: first we define the transformation from the molecular frame
to the lab frame via 

\begin{equation}
v_{i}=l_{i\alpha}v_{\alpha},\label{eq:v_i=00003Dl_ialpha v_alpha}
\end{equation}

where we used Einstein's summation convention, latin and greek indices
indicate components in the lab and molecular frame respectively, and
$l_{i\alpha}$ stands for the direction cosine between the axis $i=x^{\mathrm{L}},y^{\mathrm{L}},z^{\mathrm{L}}$
in the lab frame and the axis $\alpha=x^{\mathrm{M}},y^{\mathrm{M}},z^{\mathrm{M}}$
in the molecular frame. The direction cosines can be written in terms
of the Euler angles $\varrho\equiv\left(\alpha\beta\gamma\right)$
(see for example Sec. 2.2 in Ref. \cite{kroto_molecular_1992}).
From Sec. 4.2.5 of Ref. \cite{barron_molecular_2004} (see also \cite{andrews_1977}) we have that the isotropic orientation averages of products
of direction cosines are 

\begin{equation}
\int\mathrm{d}\varrho\,l_{i\alpha}l_{j\beta}=\frac{1}{3}\delta_{ij}\delta_{\alpha\beta},\label{eq:<ll>}
\end{equation}

\begin{equation}
\int\mathrm{d}\varrho\,l_{i\alpha}l_{j\beta}l_{k\gamma}=\frac{1}{6}\epsilon_{ijk}\epsilon_{\alpha\beta\gamma},\label{eq:<lll>}
\end{equation}

\begin{eqnarray}
\int\mathrm{d}\varrho\,l_{i\alpha}l_{j\beta}l_{k\gamma}l_{l\delta}l_{m\epsilon} & = & \frac{1}{30}\bigg\{\epsilon_{ijk}\delta_{lm}\epsilon_{\alpha\beta\gamma}\delta_{\delta\epsilon}+\epsilon_{ijl}\delta_{km}\epsilon_{\alpha\beta\delta}\delta_{\gamma\epsilon}\nonumber \\
 &  & +\epsilon_{ijm}\delta_{kl}\epsilon_{\alpha\beta\epsilon}\delta_{\gamma\delta}+\epsilon_{ikl}\delta_{jm}\epsilon_{\alpha\gamma\delta}\delta_{\beta\epsilon}\nonumber \\
 &  & +\epsilon_{ikm}\delta_{jl}\epsilon_{\alpha\gamma\epsilon}\delta_{\beta\delta}+\epsilon_{ilm}\delta_{jk}\epsilon_{\alpha\delta\epsilon}\delta_{\beta\gamma}\nonumber \\
 &  & +\epsilon_{jkl}\delta_{im}\epsilon_{\beta\gamma\delta}\delta_{\alpha\epsilon}+\epsilon_{jkm}\delta_{il}\epsilon_{\beta\gamma\epsilon}\delta_{\alpha\delta}\nonumber \\
 &  & +\epsilon_{jlm}\delta_{ik}\epsilon_{\beta\delta\epsilon}\delta_{\alpha\gamma}+\epsilon_{klm}\delta_{ij}\epsilon_{\gamma\delta\epsilon}\delta_{\alpha\beta}\bigg\}\label{eq:<lllll>}
\end{eqnarray}

where $\int\mathrm{d}\varrho\equiv\frac{1}{8\pi^{2}}\int_{0}^{2\pi}\mathrm{d}\alpha\int_{0}^{\pi}\mathrm{d}\beta\int_{0}^{2\pi}\mathrm{d}\gamma\sin\beta$.
Straightforward application of formulas \eqref{eq:<ll>}, \eqref{eq:<lll>}, and \eqref{eq:<lllll>}
yields the vector identities

\begin{equation}
\int\mathrm{d}\varrho\,\left(\vec{a}^{\mathrm{L}}\t\vec{v}^{\mathrm{L}}\right)\vec{b}^{\mathrm{L}}=\frac{1}{3}\left(\vec{a}^{\mathrm{M}}\t\vec{b}^{\mathrm{M}}\right)\vec{v}^{\mathrm{L}},\label{eq:(a.b)v}
\end{equation}

\begin{equation}
\int\mathrm{d}\varrho\,\left(\vec{a}^{\mathrm{L}}\t\vec{u}^{\mathrm{L}}\right)\left(\vec{b}^{\mathrm{L}}\t\vec{v}^{\mathrm{L}}\right)\vec{c}^{\mathrm{L}}=\frac{1}{6}\left[\left(\vec{a}^{\mathrm{M}}\x\vec{b}^{\mathrm{M}}\right)\t\vec{c}^{\mathrm{M}}\right]\left(\vec{u}^{\mathrm{L}}\x\vec{v}^{\mathrm{L}}\right),\label{eq:=00005B(axb).c=00005D(uxv)}
\end{equation}

\begin{eqnarray}
 &  & \int\mathrm{d}\varrho\,\left(\vec{a}^{\mathrm{L}}\t\vec{u}^{\mathrm{L}}\right)\left(\vec{b}^{\mathrm{L}}\t\vec{v}^{\mathrm{L}}\right)\left(\vec{c}^{\mathrm{L}}\t\vec{w}^{\mathrm{L}}\right)\left(\vec{d}^{\mathrm{L}}\t\vec{x}^{\mathrm{L}}\right)\vec{e}^{\mathrm{L}}\nonumber \\
 & = & \frac{1}{30}\bigg\{\left[\left(\vec{a}^{\mathrm{M}}\x\vec{b}^{\mathrm{M}}\right)\t\vec{c}^{\mathrm{M}}\right]\left(\vec{d}^{\mathrm{M}}\t\vec{e}^{\mathrm{M}}\right)\left[\left(\vec{u}^{\mathrm{L}}\x\vec{v}^{\mathrm{L}}\right)\t\vec{w}^{\mathrm{L}}\right]\vec{x}^{\mathrm{L}}\nonumber \\
 &  & +\left[\left(\vec{a}^{\mathrm{M}}\x\vec{b}^{\mathrm{M}}\right)\t\vec{d}^{\mathrm{M}}\right]\left(\vec{c}^{\mathrm{M}}\t\vec{e}^{\mathrm{M}}\right)\left[\left(\vec{u}^{\mathrm{L}}\x\vec{v}^{\mathrm{L}}\right)\t\vec{x}^{\mathrm{L}}\right]\vec{w}^{\mathrm{L}}\nonumber \\
 &  & +\left[\left(\vec{a}^{\mathrm{M}}\x\vec{b}^{\mathrm{M}}\right)\t\vec{e}^{\mathrm{M}}\right]\left(\vec{c}^{\mathrm{M}}\t\vec{d}^{\mathrm{M}}\right)\left(\vec{u}^{\mathrm{L}}\x\vec{v}^{\mathrm{L}}\right)\left(\vec{w}^{\mathrm{L}}\t\vec{x}^{\mathrm{L}}\right)\nonumber \\
 &  & +\left[\left(\vec{a}^{\mathrm{M}}\x\vec{c}^{\mathrm{M}}\right)\t\vec{d}^{\mathrm{M}}\right]\left(\vec{b}^{\mathrm{M}}\t\vec{e}^{\mathrm{M}}\right)\left[\left(\vec{u}^{\mathrm{L}}\x\vec{w}^{\mathrm{L}}\right)\t\vec{x}^{\mathrm{L}}\right]\vec{v}^{\mathrm{L}}\nonumber \\
 &  & +\left[\left(\vec{a}^{\mathrm{M}}\x\vec{c}^{\mathrm{M}}\right)\t\vec{e}^{\mathrm{M}}\right]\left(\vec{b}^{\mathrm{M}}\t\vec{d}^{\mathrm{M}}\right)\left(\vec{u}^{\mathrm{L}}\x\vec{w}^{\mathrm{L}}\right)\left(\vec{v}^{\mathrm{L}}\t\vec{x}^{\mathrm{L}}\right)\nonumber \\
 &  & +\left[\left(\vec{a}^{\mathrm{M}}\x\vec{d}^{\mathrm{M}}\right)\t\vec{e}^{\mathrm{M}}\right]\left(\vec{b}^{\mathrm{M}}\t\vec{c}^{\mathrm{M}}\right)\left(\vec{u}^{\mathrm{L}}\x\vec{x}^{\mathrm{L}}\right)\left(\vec{v}^{\mathrm{L}}\t\vec{w}^{\mathrm{L}}\right)\nonumber \\
 &  & +\left[\left(\vec{b}^{\mathrm{M}}\x\vec{c}^{\mathrm{M}}\right)\t\vec{d}^{\mathrm{M}}\right]\left(\vec{a}^{\mathrm{M}}\t\vec{e}^{\mathrm{M}}\right)\left[\left(\vec{v}^{\mathrm{L}}\x\vec{w}^{\mathrm{L}}\right)\t\vec{x}^{\mathrm{L}}\right]\vec{u}^{\mathrm{L}}\nonumber \\
 &  & +\left[\left(\vec{b}^{\mathrm{M}}\x\vec{c}^{\mathrm{M}}\right)\t\vec{e}^{\mathrm{M}}\right]\left(\vec{a}^{\mathrm{M}}\t\vec{d}^{\mathrm{M}}\right)\left(\vec{v}^{\mathrm{L}}\x\vec{w}^{\mathrm{L}}\right)\left(\vec{u}^{\mathrm{L}}\t\vec{x}^{\mathrm{L}}\right)\nonumber \\
 &  & +\left[\left(\vec{b}^{\mathrm{M}}\x\vec{d}^{\mathrm{M}}\right)\t\vec{e}^{\mathrm{M}}\right]\left(\vec{a}^{\mathrm{M}}\t\vec{c}^{\mathrm{M}}\right)\left(\vec{v}^{\mathrm{L}}\x\vec{x}^{\mathrm{L}}\right)\left(\vec{u}^{\mathrm{L}}\t\vec{w}^{\mathrm{L}}\right)\nonumber \\
 &  & +\left[\left(\vec{c}^{\mathrm{M}}\x\vec{d}^{\mathrm{M}}\right)\t\vec{e}^{\mathrm{M}}\right]\left(\vec{a}^{\mathrm{M}}\t\vec{b}^{\mathrm{M}}\right)\left(\vec{w}^{\mathrm{L}}\x\vec{x}^{\mathrm{L}}\right)\left(\vec{u}^{\mathrm{L}}\t\vec{v}^{\mathrm{L}}\right)\bigg\}\label{eq:5th_rank_tensor_invariant_vector_identity}
\end{eqnarray}

for arbitrary vectors $\vec{a}$, $\vec{b}$, $\vec{c}$, $\vec{d}$,
$\vec{e},$ $\vec{u}$, $\vec{v}$, $\vec{w}$, and $\vec{x}$, respectively.

\subsection{Quantum orientation averaging\label{sub:Appendix_QuantumOrientationAveraging}}
In this appendix we will derive the identities

\begin{equation}
\sum_{M_{i}}\vec{A}_{i,i}=0,\label{eq:sum_A_ii}
\end{equation}

\begin{equation}
\sum_{M_{i},M_{j}}\left(\vec{A}_{i,j}\t\vec{u}\right)\vec{A}_{j,i}=\frac{1}{3}\sum_{M_{i},M_{j}}\left(\vec{A}_{i,j}\t\vec{A}_{j,i}\right)\vec{u},\label{eq:(A.u)A=00003D(A.A)u/3}
\end{equation}

\begin{equation}
\sum_{M_{i},M_{j},M_{k}}\left(\vec{A}_{i,j}\t\vec{u}\right)\left(\vec{B}_{k,i,}\t\vec{v}\right)\vec{C}_{j,k}=\frac{1}{6}\sum_{M_{i},M_{j},M_{k}}\left[\left(\vec{A}_{i,j}\x\vec{B}_{k,i}\right)\t\vec{C}_{j,k}\right]\left(\vec{u}\x\vec{v}\right),\label{eq:(A.u)(B.v)C =00003D =00005B(AxB).C=00005D=00005Buxv=00005D/6}
\end{equation}

where $\hat{\vec{A}}$, $\hat{\vec{B}}$, and $\hat{\vec{C}}$, are
vector operators, $\vec{u}$ and $\vec{v}$ are vectors, and we use
the shorthand notation $\vec{A}_{i,j}=\bra{\alpha_{i}J_{i}M_{i}}\hat{\vec{A}}\ket{\alpha_{j}J_{j}M_{j}}$.
The state $\ket{\alpha JM}$ is an eigenfunction of the total angular
momentum operator $\hat{J}^{2}$ and of its $z$ component $\hat{J}_{z}$,
with eigenvalues $J\left(J+1\right)$ and $M$ respectively. The label
$\alpha$ indicates all the other quantum numbers required to describe
the state. 

These equations can be used to carry out the orientation averaging
procedure of the expected value of the dipole in Sec. \ref{sec:microwavePXCD}. 

The first identity is rather trivial, especially in view of its classical
analogue. The second and third identities are the quantum analogues
of Eqs. \eqref{eq:(a.b)v} and \eqref{eq:=00005B(axb).c=00005D(uxv)}
respectively. The proofs below are valid both for integer and half-integer
$J$. 

Before going into the derivation we will briefly remind the reader
of a few formulas that we will use throughout our derivation. The
spherical components of a vector are defined by (see Eq. 4.10 in \cite{brink_angular_1968})

\begin{equation}
v_{0}=v_{z},\qquad v_{\pm}=\mp\frac{1}{\sqrt{2}}\left(v_{x}\pm\i v_{y}\right).
\end{equation}

From this definition it follows that the dot product, the cross product,
and the scalar triple product can be written in terms of their spherical
components as follows:

\begin{eqnarray}
\vec{u}\t\vec{v} & = & \sum_{q=-1}^{1}\left(-1\right)^{q}u_{-q}v_{q}\label{eq:dot_product_spherical}
\end{eqnarray}

\begin{equation}
\left(\vec{u}\x\vec{v}\right)_{p}=\left(-1\right)^{p}\i\sum_{q,r=-1}^{1}\epsilon_{pqr}u_{-q}v_{-r}\label{eq:cross_product_spherical}
\end{equation}

\begin{equation}
\left(\vec{u}\x\vec{v}\right)\t\vec{w}=-\i\sum_{p,q,r=-1}^{1}\epsilon_{pqr}u_{p}v_{q}w_{r}\label{eq:scalar_triple_product_spherical}
\end{equation}

where $\epsilon_{pqr}$ is the Levi-Civita tensor for the set $\left\{ -1,0,1\right\} $
such that $\epsilon_{-1,0,1}=\epsilon_{0,1,-1}=\epsilon_{1,-1,0}=1$
and $\epsilon_{1,0,-1}=\epsilon_{-1,1,0}=\epsilon_{0,-1,1}=-1$, and
every other component is equal to zero. Note also that

\begin{equation}
\frac{1}{\sqrt{6}}\left(\begin{array}{ccc}
1 & 1 & 1\\
-1 & 0 & 1
\end{array}\right)=1,
\end{equation}

which along with the symmetry properties of the 3-j symbol for column
permutations implies that (see also Sec. 3.2 in \cite{brink_angular_1968})

\begin{equation}
\sqrt{6}\left(\begin{array}{ccc}
1 & 1 & 1\\
p & q & r
\end{array}\right)=\epsilon_{pqr}.\label{eq:Levi_Civita_3j}
\end{equation}

Another special value of the 3-j symbol is obtained by considering
the coupling to zero angular momentum $\braket{JM;00}{JM}=1$ and
the relationship between the Clebsch-Gordan coefficient and the 3-j symbol,
which yields

\begin{equation}
\left(\begin{array}{ccc}
J & 0 & J\\
-M & 0 & M
\end{array}\right)=\frac{\left(-1\right)^{J-M}}{\sqrt{2J+1}}.\label{eq:coupling_to_zero_3j}
\end{equation}

We will also use the formula (see Eq. 7.35 of \cite{brink_angular_1968}\footnote{There is a misprint in the reference.})

\begin{multline}
\sum_{\delta\epsilon\phi}\left(-1\right)^{d-\delta+e-\epsilon+f-\phi}\left(\begin{array}{ccc}
d & e & c\\
-\delta & \epsilon & \gamma
\end{array}\right)\left(\begin{array}{ccc}
e & f & a\\
-\epsilon & \phi & \alpha
\end{array}\right)\left(\begin{array}{ccc}
f & d & b\\
-\phi & \delta & \beta
\end{array}\right)\\
=\left\{ \begin{array}{ccc}
a & b & c\\
d & e & f
\end{array}\right\} \left(\begin{array}{ccc}
a & b & c\\
\alpha & \beta & \gamma
\end{array}\right),\label{eq:contraction_of_3_3j}
\end{multline}

where the symbol in curly brackets is a 6-j symbol. 

Finally, the Wigner-Eckart theorem for the spherical component $q$
of a rank $k$ tensor reads as\footnote{Our reduced matrix element contains an extra factor of $\sqrt{2J+1}$
in comparison to that defined in Ref. \cite{brink_angular_1968}.} (see \cite{brink_angular_1968})

\begin{eqnarray}
\bra{\alpha JM}T_{q}^{k}\ket{\alpha^{\prime}J^{\prime}M^{\prime}} & = & \langle\alpha J\|\boldsymbol{T}_{k}\|\alpha^{\prime}J^{\prime}\rangle\left(-1\right)^{J-M}\left(\begin{array}{ccc}
J & k & J^{\prime}\\
-M & q & M^{\prime}
\end{array}\right).\label{eq:Wigner-Eckart}
\end{eqnarray}

Now we begin with the proof of Eq. \eqref{eq:sum_A_ii}. For this case
we will drop the index $i$ on the quantum numbers and let $\alpha\neq\alpha^{\prime}$.
On the left hand side of Eq. \eqref{eq:sum_A_ii} the addends read
as

\begin{equation}
\vec{A}_{i,i}=\bra{\alpha JM}\vec{A}\ket{\alpha^{\prime}J_{}M_{}}=\langle\alpha J\|\boldsymbol{A}\|\alpha^{\prime}J\rangle\left(-1\right)^{J-M}\sum_{q}\left(\begin{array}{ccc}
J & 1 & J\\
-M & q & M
\end{array}\right)\hat{e}_q,\label{eq:A_ii}
\end{equation}

and the corresponding sum over $M$ yields

\begin{eqnarray}
\sum_{M}\left(-1\right)^{J-M}\left(\begin{array}{ccc}
J & 1 & J\\
-M & q & M
\end{array}\right)&=&\sqrt{2J+1}\sum_{M}\left(\begin{array}{ccc}
J & 0 & J\\
-M & 0 & M
\end{array}\right)\left(\begin{array}{ccc}
J & 1 & J\\
-M & q & M
\end{array}\right),\nonumber\\
&=&\sqrt{2J+1}\sum_{M,M^{\prime}}\left(\begin{array}{ccc}
J & 0 & J\\
-M & 0 & M^{\prime}
\end{array}\right)\left(\begin{array}{ccc}
J & 1 & J\\
-M & q & M^{\prime}
\end{array}\right),\nonumber\\
&=&0, \label{eq:A_ii_sumM}
\end{eqnarray}

where we used Eqs. \eqref{eq:coupling_to_zero_3j}, the selection rule $-M+M^\prime = 0$, and the orthogonality
of the 3-j symbols. Eqs. \eqref{eq:A_ii} and \eqref{eq:A_ii_sumM}
yield the first identity {[}Eq. \eqref{eq:sum_A_ii}{]}.

For the second identity, we can use Eqs. \eqref{eq:dot_product_spherical} and \eqref{eq:Wigner-Eckart} to write the addends on the left hand side of Eq. \eqref{eq:(A.u)A=00003D(A.A)u/3} as

\begin{eqnarray}
\left(\vec{A}_{i,j}\t\vec{u}\right)\vec{A}_{j,i} & = & \sum_{q,p}\left(-1\right)^{q}\bra{\alpha_{i}J_{i}M_{i}}A_{-q}\ket{\alpha_{j}J_{j}M_{j}}u_{q}\bra{\alpha_{j}J_{j}M_{j}}A_{p}\ket{\alpha_{i}J_{i}M_{i}}\hat{e}_{p},\nonumber \\
 & = & \sum_{q,p}\left(-1\right)^{q}\langle\alpha_{i}J_{i}\|\boldsymbol{A}\|\alpha_{j}J_{j}\rangle\left(-1\right)^{J_{i}-M_{i}}\left(\begin{array}{ccc}
J_{i} & 1 & J_{j}\\
-M_{i} & -q & M_{j}
\end{array}\right)u_{q}\nonumber \\
 &  & \times\langle\alpha_{j}J_{j}\|\boldsymbol{A}\|\alpha_{i}J_{i}\rangle\left(-1\right)^{J_{j}-M_{j}}\left(\begin{array}{ccc}
J_{j} & 1 & J_{i}\\
-M_{j} & p & M_{i}
\end{array}\right)\hat{e}_{p},\label{eq:(A.u)A}
\end{eqnarray}

and the corresponding sum over $M_{i}$ and $M_{j}$ yields

\begin{eqnarray}
 &  & \sum_{M_{i},M_{j}}\left(-1\right)^{J_{i}-M_{i}+J_{j}-M_{j}}\left(\begin{array}{ccc}
J_{i} & 1 & J_{j}\\
-M_{i} & -q & M_{j}
\end{array}\right)\left(\begin{array}{ccc}
J_{j} & 1 & J_{i}\\
-M_{j} & p & M_{i}
\end{array}\right)\nonumber \\
 & = & \sum_{M_{i},M_{j}}\left(-1\right)^{J_{i}-M_{i}+J_{j}-M_{j}}\left(\begin{array}{ccc}
J_{i} & J_{j} & 1\\
-M_{i} & M_{j} & -q
\end{array}\right)\left(\begin{array}{ccc}
J_{i} & J_{j} & 1\\
-M_{i} & M_{j} & -p
\end{array}\right),\nonumber\\
 & = & \left(-1\right)^{-J_{i}+J_{j}-q}\sum_{M_{i},M_{j}}\left(\begin{array}{ccc}
J_{i} & J_{j} & 1\\
-M_{i} & M_{j} & -q
\end{array}\right)\left(\begin{array}{ccc}
J_{i} & J_{j} & 1\\
-M_{i} & M_{j} & -p
\end{array}\right),\nonumber \\
 & = & \frac{\left(-1\right)^{-J_{i}+J_{j}-q}}{\sqrt{3}}\delta_{p,q},\label{eq:sum_M (A.u)A}
\end{eqnarray}

where we used the symmetry property for column exchange and for negating all $M$'s of the 3-j
symbol, the selection rule for the $M$'s to write $M_{j}=M_{i}+q$
in the exponent of $\left(-1\right)$, and the fact that $J_i + J_j + 1$ is an integer. Then we replaced $\left(-1\right)^{2M_{i}}$
by $\left(-1\right)^{2J_{i}}$, and used the orthogonality relation
of 3-j symbols. Replacing Eqs. \eqref{eq:(A.u)A} and \eqref{eq:sum_M (A.u)A}
on the left hand side of \eqref{eq:(A.u)A=00003D(A.A)u/3} we get

\begin{eqnarray}
\sum_{M_{i},M_{j}}\left(\vec{A}_{i,j}\t\vec{u}\right)\vec{A}_{j,i} & = & F\sum_{q,p}u_{q}\delta_{p,q}\hat{e}_{p}\nonumber \\
 & = & F\vec{u}\label{eq:(A.u)A=00003D(A.A)u LHS}
\end{eqnarray}

where we defined

\begin{equation}
F\equiv\frac{\left(-1\right)^{J_{j}-J_{i}}}{\sqrt{3}}\left|\langle\alpha_{i}J_{i}\|\boldsymbol{A}\|\alpha_{j}J_{j}\rangle\right|^{2}.
\end{equation}
Using Eq. \eqref{eq:sum_M (A.u)A} with $p=q$, the right hand side
of Eq. \eqref{eq:(A.u)A=00003D(A.A)u/3} yields

\begin{eqnarray}
\sum_{M_{i},M_{j}}\left(\vec{A}_{i,j}\t\vec{A}_{j,i}\right) & = & \sum_{M_{i},M_{j},q}\left(-1\right)^{q}\bra{\alpha_{i}J_{i}M_{i}}A_{-q}\ket{\alpha_{j}J_{j}M_{j}}\bra{\alpha_{j}J_{j}M_{j}}A_{q}\ket{\alpha_{i}J_{i}M_{i}},\nonumber \\
 & = & F\sum_{q}\delta_{q,q},\nonumber \\
 & = & 3F,\label{eq:(A.u)A=00003D(A.A)u RHS}
\end{eqnarray}

which in comparison with Eq. \eqref{eq:(A.u)A=00003D(A.A)u LHS} yields
the identity \eqref{eq:(A.u)A=00003D(A.A)u/3}.

For the third identity, we can use Eqs. \eqref{eq:dot_product_spherical} and \eqref{eq:Wigner-Eckart} to write the addends on the left hand side of Eq. Eq. \eqref{eq:(A.u)(B.v)C =00003D =00005B(AxB).C=00005D=00005Buxv=00005D/6} as

\begin{eqnarray}
&& \left(\vec{A}_{i,j}\t\vec{u}\right)\left(\vec{B}_{k,i,}\t\vec{v}\right)\vec{C}_{j,k}\nonumber\\
& = & \sum_{p,q,r}\left(-1\right)^{p+q}\bra{\alpha_{i}J_{i}M_{i}}A_{-p}\ket{\alpha_{j}J_{j}M_{j}}u_{p}\nonumber \\
 &  & \times\bra{\alpha_{k}J_{k}M_{k}}B_{-q}\ket{\alpha_{i}J_{i}M_{i}}v_{q}\bra{\alpha_{j}J_{j}M_{j}}C_{r}\ket{\alpha_{k}J_{k}M_{k}}\hat{e}_{r},\nonumber \\
 & = & \sum_{p,q,r}\left(-1\right)^{p+q}\langle\alpha_{i}J_{i}\|\boldsymbol{A}\|\alpha_{j}J_{j}\rangle\left(-1\right)^{J_{i}-M_{i}}\left(\begin{array}{ccc}
J_{i} & 1 & J_{j}\\
-M_{i} & -p & M_{j}
\end{array}\right)u_{p}\nonumber \\
 &  & \times\langle\alpha_{k}J_{k}\|\boldsymbol{B}\|\alpha_{i}J_{i}\rangle\left(-1\right)^{J_{k}-M_{k}}\left(\begin{array}{ccc}
J_{k} & 1 & J_{i}\\
-M_{k} & -q & M_{i}
\end{array}\right)v_{q}\nonumber \\
 &  & \times\langle\alpha_{j}J_{j}\|\boldsymbol{C}\|\alpha_{k}J_{k}\rangle\left(-1\right)^{J_{j}-M_{j}}\left(\begin{array}{ccc}
J_{j} & 1 & J_{k}\\
-M_{j} & r & M_{k}
\end{array}\right)\hat{e}_{r},\label{eq:(A.u)(B.v)C}
\end{eqnarray}

and the corresponding sum over all $M_{i}$, $M_{j}$, and $M_{k}$
yields

\begin{eqnarray}
 &  & \sum_{M_{i},M_{j},M_{k}}\left(-1\right)^{J_{i}-M_{i}+J_{j}-M_{j}+J_{k}-M_{k}}\nonumber \\
 &  & \times\left(\begin{array}{ccc}
J_{i} & 1 & J_{j}\\
-M_{i} & -p & M_{j}
\end{array}\right)\left(\begin{array}{ccc}
J_{k} & 1 & J_{i}\\
-M_{k} & -q & M_{i}
\end{array}\right)\left(\begin{array}{ccc}
J_{j} & 1 & J_{k}\\
-M_{j} & r & M_{k}
\end{array}\right),\nonumber \\
 & = & \left(-1\right)^{2J_{i}+2J_{j}+2J_{k}+3}\sum_{M_{i},M_{j},M_{k}}\left(-1\right)^{J_{i}-M_{i}+J_{j}-M_{j}+J_{k}-M_{k}}\nonumber \\
 &  & \times\left(\begin{array}{ccc}
J_{i} & J_{j} & 1\\
-M_{i} & M_{j} & -p
\end{array}\right)\left(\begin{array}{ccc}
J_{j} & J_{k} & 1\\
-M_{j} & M_{k} & r
\end{array}\right)\left(\begin{array}{ccc}
J_{k} & J_{i} & 1\\
-M_{k} & M_{i} & -q
\end{array}\right),\nonumber \\
 & = & \left(-1\right)^{2J_{k}+1}\left\{ \begin{array}{ccc}
1 & 1 & 1\\
J_{i} & J_{j} & J_{k}
\end{array}\right\} \left(\begin{array}{ccc}
1 & 1 & 1\\
r & -q & -p
\end{array}\right),\nonumber \\
 & = & \frac{\left(-1\right)^{2J_{k}+1}}{\sqrt{6}}\left\{ \begin{array}{ccc}
1 & 1 & 1\\
J_{i} & J_{j} & J_{k}
\end{array}\right\} \epsilon_{r,-q,-p},\label{eq:sum_M (A.u)(B.v)C}
\end{eqnarray}

where we used the symmetry property for column exchange of the 3-j
symbols, Eqs. \eqref{eq:Levi_Civita_3j} and \eqref{eq:contraction_of_3_3j},
and the fact that $J_{i}+J_{j}+1$ is an integer. Replacing Eqs. 
\eqref{eq:(A.u)(B.v)C} and \eqref{eq:sum_M (A.u)(B.v)C} in the left hand side of Eq. \eqref{eq:(A.u)(B.v)C =00003D =00005B(AxB).C=00005D=00005Buxv=00005D/6} and using Eq. \eqref{eq:cross_product_spherical},
we get

\begin{eqnarray}
\sum_{M_{i},M_{j},M_{k}}\left(\vec{A}_{i,j}\t\vec{u}\right)\left(\vec{B}_{k,i,}\t\vec{v}\right)\vec{C}_{j,k} & = & G\sum_{p,q,r}\left(-1\right)^{p+q}u_{p}v_{q}\epsilon_{r,-q,-p}\hat{e}_{r},\nonumber \\
 & = & \i G\sum_{p,q,r}\left(-1\right)^{r}\i\epsilon_{r,p,q}u_{-p}v_{-q}\hat{e}_{r},\nonumber \\
 & = & \i G\left(\vec{u}\x\vec{v}\right),\label{eq:(A.u)(B.v)C LHS}
\end{eqnarray}

where we defined 

\begin{equation}
G\equiv\frac{\left(-1\right)^{2J_{k}+1}}{\sqrt{6}}\langle\alpha_{i}J_{i}\|\boldsymbol{A}\|\alpha_{j}J_{j}\rangle\langle\alpha_{k}J_{k}\|\boldsymbol{B}\|\alpha_{i}J_{i}\rangle\langle\alpha_{j}J_{j}\|\boldsymbol{C}\|\alpha_{k}J_{k}\rangle\left\{ \begin{array}{ccc}
1 & 1 & 1\\
J_{i} & J_{j} & J_{k}
\end{array}\right\} .
\end{equation}

On the right hand side of the identity {[}Eq. \eqref{eq:(A.u)(B.v)C =00003D =00005B(AxB).C=00005D=00005Buxv=00005D/6}{]}
we have 

\begin{eqnarray}
\left(\vec{A}_{i,j}\x\vec{B}_{k,i}\right)\t\vec{C}_{j,k} & = & -\i\sum_{p,q,r}\epsilon_{pqr}\bra{\alpha_{i}J_{i}M_{i}}A_{p}\ket{\alpha_{j}J_{j}M_{j}}\bra{\alpha_{k}J_{k}M_{k}}B_{q}\ket{\alpha_{i}J_{i}M_{i}}\nonumber \\
 &  & \times\bra{\alpha_{j}J_{j}M_{j}}C_{r}\ket{\alpha_{k}J_{k}M_{k}},\nonumber \\
 & = & -\i\sum_{p,q,r}\epsilon_{pqr}\langle\alpha_{i}J_{i}\|\boldsymbol{A}\|\alpha_{j}J_{j}\rangle\left(-1\right)^{J_{i}-M_{i}}\left(\begin{array}{ccc}
J_{i} & 1 & J_{j}\\
-M_{i} & p & M_{j}
\end{array}\right)\nonumber \\
 &  & \times\langle\alpha_{k}J_{k}\|\boldsymbol{B}\|\alpha_{i}J_{i}\rangle\left(-1\right)^{J_{k}-M_{k}}\left(\begin{array}{ccc}
J_{k} & 1 & J_{i}\\
-M_{k} & q & M_{i}
\end{array}\right)\nonumber \\
 &  & \times\langle\alpha_{j}J_{j}\|\boldsymbol{C}\|\alpha_{k}J_{k}\rangle\left(-1\right)^{J_{j}-M_{j}}\left(\begin{array}{ccc}
J_{j} & 1 & J_{k}\\
-M_{j} & r & M_{k}
\end{array}\right),
\end{eqnarray}

and, inverting the sign of $q$ and $p$ in \eqref{eq:sum_M (A.u)(B.v)C},
the corresponding sum over $M_{i}$, $M_{j}$, and $M_{k}$ yields 

\begin{eqnarray}
\sum_{M_{i},M_{j},M_{k}}\left(\vec{A}_{i,j}\x\vec{B}_{k,i}\right)\t\vec{C}_{j,k} & = & -\i G\sum_{p,q,r}\epsilon_{pqr}\epsilon_{rqp},\nonumber \\
 & = & \i G\sum_{p,q,r}\epsilon_{pqr}^{2},\nonumber \\
 & = & 6\i G,
\end{eqnarray}

which in comparison with Eq. \eqref{eq:(A.u)(B.v)C LHS} yields Eq.
\eqref{eq:(A.u)(B.v)C =00003D =00005B(AxB).C=00005D=00005Buxv=00005D/6}.

\subsection{Transition dipoles for chiral electronic states\label{sub:AppendixDipole}}

Opposite enantiomers $R$ and $L$ are related to each other via an
inversion, therefore their bound and scattering electronic wave functions
satisfy

\begin{equation}
\psi_{R}\left(\vec{r}\right)=\psi_{L}\left(-\vec{r}\right),
\end{equation}

\begin{equation}
\psi_{\vec{k},R}\left(\vec{r}\right)=\psi_{-\vec{k},L}\left(-\vec{r}\right).\label{eq:scattering_wavefunctions_R_L}
\end{equation}

Then, for the transition dipole between two electronic bound states
$\psi$ and $\psi^{\prime}$ we have

\begin{eqnarray}
\vec{d}_{R} & \equiv & -\int\d\vec{r}\,\psi_{R}^{\prime*}\left(\vec{r}\right)\vec{r}\psi_{R}\left(\vec{r}\right),\nonumber \\
 & = & \int\d\vec{r}\,\psi_{L}^{\prime*}\left(-\vec{r}\right)\left(-\vec{r}\right)\psi_{L}\left(-\vec{r}\right),\nonumber \\
 & = & \int\d\vec{r}\,\psi_{L}^{\prime*}\left(\vec{r}\right)\vec{r}\psi_{L}\left(\vec{r}\right),\nonumber \\
 & = & -\vec{d}_{L},
\end{eqnarray}

as expected. For the transition dipole between the bound state $\psi$
and the scattering state $\psi_{\vec{k}}$ one has to be more careful
because of the vector nature of the photoelectron momentum $\vec{k}$.
In this case we have

\begin{eqnarray}
\vec{D}_{R}(\vec{k}) & = & -\int\mathrm{d}\vec{r}\,\psi_{\vec{k},R}^{*}\left(\vec{r}\right)\vec{r}\psi_{R}\left(\vec{r}\right),\nonumber \\
 & = & \int\mathrm{d}\vec{r}\,\psi_{-\vec{k},L}^{*}\left(-\vec{r}\right)\left(-\vec{r}\right)\psi_{L}\left(-\vec{r}\right),\nonumber \\
 & = & \int\mathrm{d}\vec{r}\,\psi_{-\vec{k},L}^{*}\left(\vec{r}\right)\vec{r}\psi_{L}\left(\vec{r}\right),\nonumber \\
 & = & -\vec{D}_{L}(-\vec{k}).\label{eq:D_R(k)=-D_L(-k)}
\end{eqnarray}

Using Eq. \eqref{eq:D_R(k)=-D_L(-k)} it is a simple matter to confirm that the molecular term in Eq. \eqref{eq:j_lab_total_factored} does indeed have opposite sign for opposite enantiomers:

\begin{eqnarray}
\chi_{m}^{R} & = & \frac{1}{6}\int\mathrm{d}\Omega_{k}\left[\mathrm{i}\vec{D}_{R}^{*}\left(\vec{k}\right)\x\vec{D}_{R}\left(\vec{k}\right)\right]\t\vec{k},\nonumber\\
 & = & \frac{1}{6}\int\mathrm{d}\Omega_{k}\left[\mathrm{i}\vec{D}_{L}^{*}\left(-\vec{k}\right)\x\vec{D}_{L}\left(-\vec{k}\right)\right]\t\vec{k},\nonumber\\
 & = & -\frac{1}{6}\int\mathrm{d}\Omega_{k^{\prime}}\left[\mathrm{i}\vec{D}_{L}^{*}\left(\vec{k}^{\prime}\right)\x\vec{D}_{L}\left(\vec{k}^{\prime}\right)\right]\t\vec{k}^{\prime}\nonumber,\\
 & = & -\chi_{m}^{L},\label{eq:chi_R=-chi_L}
\end{eqnarray}

where we did the change of variable $\vec{k}^\prime = -\vec{k}$ in the third line. 

\subsection{Recovering Ritchie's formula\label{sub:AppendixRitchie}}

In Ritchie's original derivation \cite{ritchie_theory_1976} the $b_{1}$
factor is given by 

\begin{eqnarray}
b_{1} & = & \left|\tilde{\mathcal{E}}\right|^{2}\frac{\left(4\pi\right)^{2}}{3}\sum_{l_{j},m_{j},\lambda_{j},\mu_{j},m_{1},\mu_{1}}\braoket{\psi_{i}}{rY_{1\mu_{1}}^{*}}{\psi_{\lambda_{j},\mu_{j}}^{\left(-\right)}}\braoket{\psi_{l_{j}m_{j}}^{\left(-\right)}}{rY_{1m_{1}}}{\psi_{i}} \nonumber\\
&& \times \left(-1\right)^{1+m_{1}+m_{j}}3\sqrt{\left(2l_{j}+1\right)\left(2\lambda_{j}+1\right)}\nonumber \\
 &  & \times\left(\begin{array}{ccc}
l_{j} & \lambda_{j} & 1\\
0 & 0 & 0
\end{array}\right)\left(\begin{array}{ccc}
1 & 1 & 1\\
\sigma & -\sigma & 0
\end{array}\right) \nonumber\\
& & \left(\begin{array}{ccc}
l_{j} & \lambda_{j} & 1\\
m_{j} & -\mu_{j} & -\left(m_{j}-\mu_{j}\right)
\end{array}\right)\left(\begin{array}{ccc}
1 & 1 & 1\\
m_{1} & -\mu_{1} & -\left(m_{j}-\mu_{j}\right)
\end{array}\right),\label{eq:b1_Ritchie-1}
\end{eqnarray}

where the different prefactor in comparison with Eq. (11) in \cite{ritchie_theory_1976}
is because we take $W(\vec{k}^{\mathrm{L}})=\left|\braoket{\psi_{\vec{k}}^{\left(-\right)}}{\hat{e}_{\sigma}}{\psi_{i}}\right|^{2}$
in agreement with Eqs. \eqref{eq:W(k^L)}, \eqref{eq:j^L_and_b1},
\eqref{eq:j_mol}, and \eqref{eq:j_lab_total}. If we define 

\begin{equation}
D_{q}^{l_{j}m_{j}}\equiv\sqrt{\frac{4\pi}{3}}\braoket{\psi_{l_{j}m_{j}}^{\left(-\right)}}{rY_{1q}}{\psi_{i}},
\end{equation}
use Eq. \eqref{eq:cross_product_spherical} for the cross product
in spherical components, along with the properties $\epsilon_{pqr}=-\epsilon_{-p,-q,-r}$,
$\left(v_{q}\right)^{*}=\left(-1\right)^{q}\left(\vec{v}^{*}\right)_{-q}$,
Eq. \eqref{eq:Levi_Civita_3j}, and the selection rule $m_{1}-\mu_{1}-m_{j}+\mu_{j}=0$
of the 3-j symbol, we obtain

\begin{eqnarray}
 &  & \sum_{m_{1},\mu_{1}}\left(-1\right)^{m_{1}}\braoket{\psi_{i}}{rY_{1\mu_{1}}^{*}}{\psi_{\lambda_{j}\mu_{j}}^{\left(-\right)}}\braoket{\psi_{l_{j}m_{j}}^{\left(-\right)}}{rY_{1m_{1}}}{\psi_{i}}\left(\begin{array}{ccc}
1 & 1 & 1\\
m_{1} & -\mu_{1} & -\left(m_{j}-\mu_{j}\right)
\end{array}\right),\nonumber \\
 & = & \frac{3}{4\pi}\sum_{m_{1},\mu_{1}}\left(-1\right)^{m_{1}}\left(D_{\mu_{1}}^{\lambda_{j}\mu_{j}}\right)^{*}D_{m_{1}}^{l_{j}m_{j}}\left(\begin{array}{ccc}
1 & 1 & 1\\
m_{1} & -\mu_{1} & -\left(m_{j}-\mu_{j}\right)
\end{array}\right),\nonumber \\
 & = & \frac{3}{4\pi\sqrt{6}}\sum_{m_{1},\mu_{1}}\left(-1\right)^{m_{1}-\mu_{1}}\epsilon_{m_{1},-\mu_{1},\mu_{j}-m_{j}}\left(\vec{D}^{\lambda_{j}\mu_{j}*}\right)_{-\mu_{1}}D_{m_{1}}^{l_{j}m_{j}},\nonumber \\
 & = & -\i\frac{1}{4\pi}\sqrt{\frac{3}{2}}\left(-1\right)^{m_{j}-\mu_{j}}\i\sum_{m_{1},\mu_{1}}\epsilon_{m_{j}-\mu_{j},\mu_{1},-m_{1}}\left(\vec{D}^{\lambda_{j}\mu_{j}*}\right)_{-\mu_{1}}D_{m_{1}}^{l_{j}m_{j}},\nonumber \\
 & = & -\frac{1}{4\pi}\sqrt{\frac{3}{2}}\left(\i\vec{D}^{\lambda_{j}\mu_{j}*}\x\vec{D}^{l_{j}m_{j}}\right)_{m_{j}-\mu_{j}}.\label{eq:b1_cross_product}
\end{eqnarray}

Then we can use the integral of three spherical harmonics, 

\begin{multline}
\int\mathrm{d}\Omega_{k}Y_{l_{j},m_{j}}Y_{\lambda_{j},-\mu_{j}}Y_{1,\mu_{j}-m_{j}}\\
=\sqrt{\frac{3\left(2l_{j}+1\right)\left(2\lambda_{j}+1\right)}{4\pi}}\left(\begin{array}{ccc}
l_{j} & \lambda_{j} & 1\\
0 & 0 & 0
\end{array}\right)\left(\begin{array}{ccc}
l_{j} & \lambda_{j} & 1\\
m_{j} & -\mu_{j} & \mu_{j}-m_{j}
\end{array}\right),
\end{multline}

equation \eqref{eq:dot_product_spherical} for the dot product in spherical
components, and the selection rule for the sum of the $M$'s in the 3-j symbol to obtain 

\begin{eqnarray}
 &  & -\frac{1}{4\pi}\sqrt{\frac{3}{2}}\sum_{m_{j},\mu_{j}}\left(-1\right)^{m_{j}}\left(\i\vec{D}^{\lambda_{j}\mu_{j}*}\x\vec{D}^{l_{j}m_{j}}\right)_{m_{j}-\mu_{j}}\sqrt{\left(2l_{j}+1\right)\left(2\lambda_{j}+1\right)}\nonumber\\
 & & \times \left(\begin{array}{ccc}
l_{j} & \lambda_{j} & 1\nonumber\\
0 & 0 & 0
\end{array}\right)\left(\begin{array}{ccc}
l_{j} & \lambda_{j} & 1\\
m_{j} & -\mu_{j} & -\left(m_{j}-\mu_{j}\right)
\end{array}\right)\nonumber \\
 & = & -\frac{1}{4\pi}\sqrt{\frac{3}{2}}\sqrt{\frac{4\pi}{3}}\sum_{m_{j},\mu_{j}}\left(-1\right)^{m_{j}}\left(\i\vec{D}^{\lambda_{j}\mu_{j}*}\x\vec{D}^{l_{j}m_{j}}\right)_{m_{j}-\mu_{j}}\int\mathrm{d}\Omega_{k}Y_{l_{j},m_{j}}Y_{\lambda_{j},-\mu_{j}}Y_{1,\mu_{j}-m_{j}},\nonumber \\
 & = & -\frac{1}{4\pi}\sqrt{\frac{3}{2}}\sqrt{\frac{4\pi}{3}}\sum_{m_{j},\mu_{j}}\left(-1\right)^{m_{j}-\mu_{j}}\left(\i\vec{D}^{\lambda_{j}\mu_{j}*}\x\vec{D}^{l_{j}m_{j}}\right)_{m_{j}-\mu_{j}}\int\mathrm{d}\Omega_{k}Y_{\lambda_{j},\mu_{j}}^{*}Y_{l_{j},m_{j}}Y_{1,\mu_{j}-m_{j}},\nonumber \\
 & = & -\frac{1}{4\pi}\sqrt{\frac{3}{2}}\sqrt{\frac{4\pi}{3}}\sum_{m_{j},\mu_{j},q}\left(-1\right)^{q}\left(\i\vec{D}^{\lambda_{j}\mu_{j}*}\x\vec{D}^{l_{j}m_{j}}\right)_{q}\int\mathrm{d}\Omega_{k}Y_{\lambda_{j},\mu_{j}}^{*}Y_{l_{j},m_{j}}Y_{1,-q},\nonumber \\
 & = & -\frac{1}{4\pi}\sqrt{\frac{3}{2}}\sum_{m_{j},\mu_{j}}\int\mathrm{d}\Omega_{k}Y_{\lambda_{j},\mu_{j}}^{*}Y_{l_{j},m_{j}}\left[\left(\i\vec{D}^{\lambda_{j}\mu_{j}*}\x\vec{D}^{l_{j}m_{j}}\right)\t\hat{k}\right].\label{eq:b1_scalar_triple_product}
\end{eqnarray}

Finally, according to Eq. (10) in Ritchie's \cite{ritchie_theory_1976},
the scattering wave function is expanded as

\begin{equation}
\psi_{\vec{k}}^{\left(-\right)}\left(\vec{r}\right)=4\pi\sum\limits _{l_{j},m_{j}}\psi_{l_{j}m_{j}}^{\left(-\right)}\left(\vec{r}\right)Y_{l_{j}m_{j}}^{*}(\hat{k})
\end{equation}

and therefore the dipole transition vector reads as

\begin{eqnarray}
\vec{D} & = & \braoket{\psi_{\vec{k}}^{\left(-\right)}}{\vec{d}}{\psi_{i}},\nonumber \\
 & = & -4\pi\sum_{l_{j},m_{j},q}\braoket{\psi_{l_{j}m_{j}}^{\left(-\right)}}{\sqrt{\frac{4\pi}{3}}rY_{1,q}\hat{e}_{q}}{\psi_{i}}Y_{l_{j}m_{j}}(\hat{k}),\nonumber \\
 & = & -4\pi\sum_{l_{j},m_{j}}\vec{D}^{l_{j}m_{j}}Y_{l_{j}m_{j}}(\hat{k}).\label{eq:D_partial_wave_expansion}
\end{eqnarray}

Then, putting together Eqs. \eqref{eq:b1_Ritchie-1}, \eqref{eq:b1_cross_product},
\eqref{eq:b1_scalar_triple_product}, and \eqref{eq:D_partial_wave_expansion},
and using Eqs. \eqref{eq:j_lab_total_factored}  and \eqref{eq:Levi_Civita_3j},
we obtain

\begin{eqnarray}
b_{1} & = & \left|\mathcal{\tilde{E}}\right|^{2}\left(4\pi\right)^{2}\left(\begin{array}{ccc}
1 & 1 & 1\\
\sigma & -\sigma & 0
\end{array}\right)\frac{1}{4\pi}\sqrt{\frac{3}{2}}\nonumber\\
&&\times\sum_{l_{j},m_{j},\lambda_{j},\mu_{j}}\int\mathrm{d}\Omega_{k}Y_{\lambda_{j},\mu_{j}}^{*}Y_{l_{j},m_{j}}\left[\left(\i\vec{D}^{\lambda_{j}\mu_{j}*}\x\vec{D}^{l_{j}m_{j}}\right)\t\hat{k}\right]\nonumber\\
 & = & \frac{1}{8\pi k}\sigma\left|\mathcal{\tilde{E}}\right|^{2}\int\mathrm{d}\Omega_{k}\left[\left(\i\vec{D}^{*}\x\vec{D}\right)\t\vec{k}\right]\nonumber\\
 & = & \frac{3}{4\pi}\frac{j_{z}^{\mathrm{L}}}{k}
\end{eqnarray}

which shows that Ritchie's expression for $b_{1}$ is equivalent to
the one derived here.

\subsection{Circular pump + linear probe\label{sub:AppendixCircLin}}

In this appendix we derive Eqs. \eqref{eq:j_circ+linear} and \eqref{eq:D12_circ_linear}
from Eqs. \eqref{eq:j_diag} and \eqref{eq:j_cross} for the case
when the pump is circularly polarized according to Eq. \eqref{eq:field}
and the probe is linearly polarized along $\hat{x}^{\mathrm{L}}$.
From the selection rules already discussed in Sec. \ref{sub:Generalized PXECD}
we immediately see that
the first and last terms in $\vec{j}_{\mathrm{diag}}^{\mathrm{L}}$ [Eq. \eqref{eq:j_diag}], $\vec{j}_{\mathrm{noncopl}}^{\mathrm{L}}$ [Eq. \eqref{eq:j_noncopl}]
and the second term in $\vec{j}_{\mathrm{ellip}}^{\mathrm{L}}$ [Eq. \eqref{eq:j_ellip}] vanish.
Furthermore, the remaining terms in $\vec{j}_{\mathrm{diag}}^{\mathrm{L}}$ [Eq. \eqref{eq:j_diag}]
are purely imaginary and also vanish, which only leaves $\vec{j}_{\mathrm{lin}}^{\mathrm{L}}$ [Eq. \eqref{eq:j_lin}]
and the first term in $\vec{j}_{\mathrm{ellip}}^{\mathrm{L}}$ [Eq. \eqref{eq:j_ellip}]. Replacing
the field terms in Eq. \eqref{eq:j_ellip} we obtain

\begin{eqnarray}
\vec{j}_{\mathrm{ellip}}^{\mathrm{L}}\left(k\right) & = & \frac{\i\sigma\tilde{\mathcal{E}}}{30}\left(\vec{d}_{0,2}^{\mathrm{M}}\x\vec{d}_{1,0}^{\mathrm{M}}\right)\t\int\mathrm{d}\Omega_{k}^{\mathrm{M}}\left[\left(\vec{D}_{2}^{\mathrm{M}*}\t\vec{D}_{1}^{\mathrm{M}}\right)\vec{k}^{\mathrm{M}}\right]\e^{\i\omega_{21}\tau}\hat{z}^{\mathrm{L}}+\mathrm{c.c.},\label{eq:j_ellip_circ+x}
\end{eqnarray}

whereas for Eq. \eqref{eq:j_lin} we obtain

\begin{eqnarray}
\vec{j}_{\mathrm{lin}}^{\mathrm{L}}\left(k\right) & = & \frac{\i\sigma\tilde{\mathcal{E}}}{60}\bigg\{\int\mathrm{d}\Omega_{k}^{\mathrm{M}}\left[\left(\vec{d}_{0,2}^{\mathrm{M}}\x\vec{D}_{2}^{\mathrm{M}*}\right)\t\vec{k}^{\mathrm{M}}\right]\left(\vec{d}_{1,0}^{\mathrm{M}}\t\vec{D}_{1}^{\mathrm{M}}\right)\nonumber \\
 &  & +\int\mathrm{d}\Omega_{k}^{\mathrm{M}}\left[\left(\vec{d}_{0,2}^{\mathrm{M}}\x\vec{D}_{1}^{\mathrm{M}}\right)\t\vec{k}^{\mathrm{M}}\right]\left(\vec{D}_{2}^{\mathrm{M}*}\t\vec{d}_{1,0}^{\mathrm{M}}\right)\nonumber \\
 &  & +\int\mathrm{d}\Omega_{k}^{\mathrm{M}}\left[\left(\vec{D}_{2}^{\mathrm{M}*}\x\vec{d}_{1,0}^{\mathrm{M}}\right)\t\vec{k}^{\mathrm{M}}\right]\left(\vec{d}_{0,2}^{\mathrm{M}}\t\vec{D}_{1}^{\mathrm{M}}\right)\nonumber \\
 &  & -\int\mathrm{d}\Omega_{k}^{\mathrm{M}}\left[\left(\vec{d}_{1,0}^{\mathrm{M}}\x\vec{D}_{1}^{\mathrm{M}}\right)\t\vec{k}^{\mathrm{M}}\right]\left(\vec{d}_{0,2}^{\mathrm{M}}\t\vec{D}_{2}^{\mathrm{M}*}\right)\bigg\}\e^{\i\omega_{21}\tau}\hat{z}^{\mathrm{L}}\nonumber \\
 &  & +\mathrm{c.c.}
\end{eqnarray}

where $\tilde{\mathcal{E}}=\tilde{\mathcal{E}}_{1}^{*}\tilde{\mathcal{E}}_{2}^{*}\tilde{\mathcal{E}}_{1}\tilde{\mathcal{E}}_{2}$.
Now, in order to extract $\vec{d}_{0,2}^{\mathrm{M}}$ and $\vec{d}_{1,0}^{\mathrm{M}}$
from the integrals we begin by reordering the expression as 

\begin{eqnarray}
\vec{j}_{\mathrm{lin}}^{\mathrm{L}}\left(k\right) & = & \frac{\i\sigma\tilde{\mathcal{E}}}{60}\bigg\{\int\mathrm{d}\Omega_{k}^{\mathrm{M}}\left[\left(\vec{D}_{2}^{\mathrm{M}*}\x\vec{k}^{\mathrm{M}}\right)\t\vec{d}_{0,2}^{\mathrm{M}}\right]\left(\vec{D}_{1}^{\mathrm{M}}\t\vec{d}_{1,0}^{\mathrm{M}}\right)\nonumber \\
 &  & +\int\mathrm{d}\Omega_{k}^{\mathrm{M}}\left[\left(\vec{D}_{1}^{\mathrm{M}}\x\vec{k}^{\mathrm{M}}\right)\t\vec{d}_{0,2}^{\mathrm{M}}\right]\left(\vec{D}_{2}^{\mathrm{M}*}\t\vec{d}_{1,0}^{\mathrm{M}}\right)\nonumber \\
 &  & -\int\mathrm{d}\Omega_{k}^{\mathrm{M}}\left[\left(\vec{D}_{2}^{\mathrm{M}*}\x\vec{k}^{\mathrm{M}}\right)\t\vec{d}_{1,0}^{\mathrm{M}}\right]\left(\vec{D}_{1}^{\mathrm{M}}\t\vec{d}_{0,2}^{\mathrm{M}}\right)\nonumber \\
 &  & -\int\mathrm{d}\Omega_{k}^{\mathrm{M}}\left[\left(\vec{D}_{1}^{\mathrm{M}}\x\vec{k}^{\mathrm{M}}\right)\t\vec{d}_{1,0}^{\mathrm{M}}\right]\left(\vec{D}_{2}^{\mathrm{M}*}\t\vec{d}_{0,2}^{\mathrm{M}}\right)\bigg\}\e^{\i\omega_{21}\tau}\hat{z}^{\mathrm{L}}\nonumber \\
 &  & +\mathrm{c.c.},
\end{eqnarray}

to apply the vector identity $\left(\vec{a}\t\vec{c}\right)(\vec{b}\t\vec{d})-(\vec{a}\t\vec{d})(\vec{b}\t\vec{c})=(\vec{a}\x\vec{b})\t(\vec{c}\x\vec{d})$,
which yields

\begin{eqnarray}
\vec{j}_{\mathrm{lin}}^{\mathrm{L}}\left(k\right) & = & \frac{\i\sigma\tilde{\mathcal{E}}}{60}\left(\vec{d}_{0,2}^{\mathrm{M}}\x\vec{d}_{1,0}^{\mathrm{M}}\right)\t\bigg\{\int\mathrm{d}\Omega_{k}^{\mathrm{M}}\left[\left(\vec{D}_{2}^{\mathrm{M}*}\x\vec{k}^{\mathrm{M}}\right)\x\vec{D}_{1}^{\mathrm{M}}\right]\nonumber \\
 &  & +\int\mathrm{d}\Omega_{k}^{\mathrm{M}}\left[\left(\vec{D}_{1}^{\mathrm{M}}\x\vec{k}^{\mathrm{M}}\right)\x\vec{D}_{2}^{\mathrm{M}*}\right]\bigg\}\e^{\i\omega_{21}\tau}\hat{z}^{\mathrm{L}}\nonumber \\
 &  & +\mathrm{c.c.}
\end{eqnarray}

Now we use the vector identity $(\vec{a}\x\vec{b})\x\vec{c}=(\vec{a}\t\vec{c})\vec{b}-(\vec{b}\t\vec{c})\vec{a}$
to get

\begin{eqnarray}
\vec{j}_{\mathrm{lin}}^{\mathrm{L}}\left(k\right) & = & \frac{\i\sigma\tilde{\mathcal{E}}}{60}\left(\vec{d}_{0,2}^{\mathrm{M}}\x\vec{d}_{1,0}^{\mathrm{M}}\right)\int\mathrm{d}\Omega_{k}^{\mathrm{M}}\bigg\{2\left(\vec{D}_{2}^{\mathrm{M}*}\t\vec{D}_{1}^{\mathrm{M}}\right)\vec{k}^{\mathrm{M}}-\left(\vec{k}^{\mathrm{M}}\t\vec{D}_{1}^{\mathrm{M}}\right)\vec{D}_{2}^{\mathrm{M}*}\nonumber \\
 &  & -\left(\vec{k}^{\mathrm{M}}\t\vec{D}_{2}^{\mathrm{M}*}\right)\vec{D}_{1}^{\mathrm{M}}\bigg\}\e^{\i\omega_{21}\tau}\hat{z}^{\mathrm{L}}+\mathrm{c.c.}\label{eq:j_lin_circ+x}
\end{eqnarray}

Adding Eqs. \eqref{eq:j_ellip_circ+x} and \eqref{eq:j_lin_circ+x}
yields Eqs. \eqref{eq:j_circ+linear} and \eqref{eq:D12_circ_linear}.

\bibliographystyle{apsrev4-1}
\bibliography{MyLibrary}

\end{document}